\documentclass[12pt,onecolumn, floatfix, altaffilletter, superscriptaddress,tightenlines, showpacs, showkeys, preprintnumbers, nofootinbib]{revtex4-1}
\pdfoutput=1
\usepackage[colorlinks=true,citecolor=blue,linkcolor=blue,breaklinks=true]{hyperref}
\usepackage{amsmath,amssymb}
\usepackage{epsfig} 
\usepackage{graphicx}        
\usepackage{url}
\usepackage{color}
\usepackage{multirow}
\usepackage{placeins}
\usepackage[dvipsnames]{xcolor}
\usepackage{braket}
\usepackage{float}
\hypersetup{
  colorlinks=true,
  linkcolor=red,
  filecolor=magenta,   
  urlcolor=blue,
  citecolor=blue,
} 
\providecommand{\xlink}[1]
 {\href{http://arxiv.org/abs/#1}{arXiv:#1}}
 
\allowdisplaybreaks

\bibpunct{[}{]}{,}{n}{}{,}

\def\beq{\begin{equation}}
\def\eeq{\end{equation}}
\def\bea{\begin{eqnarray}}
\def\eea{\end{eqnarray}}
\makeatletter
\@addtoreset{equation}{section}
\renewcommand{\theequation}{\thesection.\arabic{equation}}

\def\thesection{\Roman{section}}

\begin{document}

\title{Non-thermal production of lepton asymmetry and dark matter in minimal seesaw with right handed
neutrino induced Higgs potential}
\author{Rome Samanta}
\email{romesamanta@gmail.com}
\affiliation{Physics and Astronomy, University of Southampton, Southampton, SO17 1BJ, U.K.}
\author{Anirban Biswas}
\email{anirban.biswas.sinp@gmail.com}
\affiliation{School of Physical Sciences, Indian Association for the Cultivation of Science,
2A $\&$ 2B Raja S.C. Mullick Road, Kolkata 700032, India}

\author{Sukannya Bhattacharya}
\email{sukannya@prl.res.in }
\affiliation{Theoretical Physics Division, Physical Research Laboratory, Navrangpura, Ahmedabad-380009, India}

\begin{abstract} 
Within Type-I seesaw mechanism, Higgs mass can be dynamically generated via quantum effects of the right handed  neutrinos assuming the potential is nearly conformal at the Ultra-Violet. The scenario, named as the ``Neutrino Option" allows RH neutrino mass scale upto $M \lesssim$ $10^7$ GeV  to be consistent with light neutrino masses, mixing and Higgs mass. Therefore, it is not consistent with standard hierarchical thermal leptogenesis. Parameter space for thermal  resonant leptogenesis is highly constrained in this model. We point out that non-thermal pair production of RH neutrinos from inflaton decay corresponds in general to a mild degree of resonance in the CP asymmetry parameter and allows RH mass scale to be smaller more than by an order of magnitude than the thermal strong resonance case. Within the similar parameter space of thermal leptogenesis, RH neutrinos can also be produced from inflaton decay along with a Dark Matter having  mass $M_{\rm DM}\lesssim$ 320 MeV.  The main constraint in the latter scenario comes from the Ly$\alpha$ constraints on Dark Matter free streaming. We show in addition, that the Neutrino Option introduces a `phantom window' for the RH mass scale, in which contrary to the usual scenarios,  CP asymmetry parameter for leptogenesis decreases with the increase of the RH mass scale and minimally fine-tuned  seesaw models naturally exhibit this `phantom window'.  \\


\end{abstract}

\maketitle

\section{Introduction}
The fact that light neutrinos have mass is well established now, thanks to dedicated neutrino oscillation experiments\cite{pdg,t2k,t2k1,t2k2,nova1,nova2,minos,dayabay,reno,dc,ic,sk,globalfit}. The simplest and minimal theory which explains light neutrino masses beyond the Standard Model (SM) is the Type-I seesaw mechanism\cite{sw1,sw2,sw3}. In this mechanism, minimum requirement of two heavy right handed (RH) neutrinos  to generate light Majorana neutrino states also facilitates  lepton number violating processes in the early universe which may be the underlying theory behind the observed dominance of matter over antimatter\cite{planck}. A lepton asymmetry (leptogenesis) which can further be converted into baryon asymmetry (baryogenesis) by $ B-L$ - conserving sphaleron processes\cite{sph1,sph2} may originate from several sources, e.g., primordial gravitational waves\cite{gravlep1,gravlep2,gravlep3}, due to asymmetric propagation of leptons and anti-lepton in curved space time\cite{curved1,curved2}, CP violating couplings between the Ricci scalar and fermions \cite{rici1,rici2,rici3,rici4} etc.. However, in the context of particle physics, leptogenesis mechanisms involving RH neutrinos are widely studied since they are automatic consequences of the RH neutrino-extended theories of SM which also explain non-vanishing light neutrino masses. There are several variations of  RH neutrino-induced leptogenesis mechanisms. Leptogenesis from RH neutrino decays\cite{fuku,pilaf,bari,nir,rio}, leptogenesis from RH neutrino oscillation\cite{Ars} and a recently proposed mechanism of leptogenesis from Higgs decays\cite{Ham}. In this work we stick to the simplest one, wherein CP violating and out of equilibrium decays of RH neutrinos create lepton asymmetry. It can be shown that if the RH neutrino masses are hierarchical, the minimum RH mass scale required for successful leptogenesis is $M_1\gtrsim10^9$ GeV- known as Davidson-Ibarra (DI) bound\cite{ibarra}. However, due to the RH neutrino involved Yukawa interactions, the {\it classical} Higgs potential suffers radiative corrections which monotonically increase with the increase of the RH masses. Not allowing the correction to the Higgs mass to exceed more than $\mathcal{O}{\rm (TeV)}$-so called the naturalness problem\cite{n1,n2,n3}, puts an upper bound on the RH mass scale $M\lesssim 10^7$ GeV which is well below the standard DI bound. Therefore, one needs to go beyond the hierarchical leptogenesis. One such mechanism is resonant-leptogenesis\cite{pilaf,dev1,dev2} where due to the presence of two strongly quasi-degenerate RH neutrinos, the CP asymmetry parameter responsible for leptogenesis gets resonantly enhanced and opens up  possibilities for successful leptogenesis even at RH mass scale $\mathcal{O}{\rm (TeV)}$.

Recently, it has been proposed (the idea named as ``Neutrino Option") that the mentioned naturalness problem can be avoided if the Higgs mass itself is generated due to the radiative corrections induced by the RH neutrinos assuming the tree-level potential is nearly conformal at the Ultra-Violet (UV)\cite{Neo1,Neo2}. The RH neutrino masses which break the conformal symmetry can be dynamically generated\cite{Neoconf} which then trigger the Electroweak symmetry breaking. However,  to be consistent with the Higgs mass and the light neutrino masses, the maximum RH mass scale one can reach up to is again $M\lesssim 10^7$ GeV\cite{lepNeo1,lepNeo2} and thus one needs to consider leptogenesis due to quasi-degenerate heavy neutrinos. Interestingly, unlike the two RH neutrino seesaw model subjected to naturalness constraints, now one can not lower the mass scale up to TeV\cite{Natlep} to have successful leptogenesis. The parameter space is highly constrained, e.g., for normal light neutrino mass ordering one obtains  $M\lesssim 1.2\times 10^6$ GeV\cite{lepNeo1}. The primary restriction comes from the thermal production of the RH neutrinos through Yukawa coupling and consequently a washout of the produced lepton asymmetry. The strength of the washout increases with the decrease of the RH mass scale due to the conditions imposed by the Neutrino Option (NeO) and thus even with a strong resonant enhancement ($\Gamma_i\simeq \Delta M$\cite{pilaf}, where $\Gamma_i$ is the decay width of the $i$th heavy neutrino and $\Delta M=M_2-M_1$) in the CP asymmetry parameter, one cannot lower the mass scale beyond $\sim 10^6$ GeV.

 In this work, in search for a larger parameter space for leptogenesis within NeO, we consider an alternative production mechanism for the RH neutrinos. We first sacrifice the fact, that the RH neutrinos are produced via Yukawa coupling by assuming $T_{\rm RH}<M_i$, where $T_{\rm RH}$ is the reheat temperature after inflation. Then we consider that RH neutrinos get produced non-thermally by inflaton decay which is a standard practice in the studies of non-thermal leptogenesis\cite{nth1,nth2,nth3,nth4,nth5,devnth,nth6,nth7,Borah:2020wyc}. Once we have $T_{\rm RH}<M_i$\footnote{In the numerical computation this condition has been implemented properly, using exact value of the temperature where the washout processes go out of equilibrium. }, the lepton asymmetry washout processes can be neglected\cite{bari,nir} because in that case the thermal bath does not have sufficient energy to reproduce RH neutrinos. Thus the final lepton asymmetry depends on the inflaton mass ($m_\phi$), $T_{\rm RH}$ and the branching ratio for inflaton to RH neutrino decays. Without going into the details of a complete inflationary dynamics which includes the process of reheating, we consider instantaneous reheating\cite{nth1,devnth,rh1} after inflation.  We show that in the case of pair production of the RH neutrinos from inflaton decay,  depending on the $T_{\rm RH}$ and inflaton mass, lower bounds on  the RH mass scale  can be relaxed by more than an order of magnitude than what is obtained in the thermal case. Allowed parameter space exhibits in general mild-resonant ($\Gamma_i \ll \Delta M$) solutions\footnote{Whenever we mention mild-resonant solutions, we always mean the departure ($\Gamma_i \ll \Delta M$) from the Pilaftsis-Underwood strong resonant condition $\Gamma_i\simeq \Delta M$\cite{pilaf}.}. Then we show that the non-thermal scenario  also allows simultaneous   production of RH neutrinos and Dark Matter (DM) almost within the same allowed range of RH neutrino masses, obtained in the case of successful  thermal leptogenesis. 
The non-thermal DM produced through freeze-in is also another interesting proposal like non-thermal leptogenesis, explaining the null results in direct detection experiments naturally \cite{Aprile:2018dbl} and has been
extensively studied in last few years \cite{Hall:2009bx, Biswas:2015sva, Konig:2016dzg, Biswas:2016bfo,
Biswas:2016iyh, Biswas:2018aib, Biswas:2019iqm}. 
In the present work, one of the important constraints comes from the Lyman-$\alpha$
forest data on DM free streaming ($\lambda_{FS}\lesssim$ 1 Mpc, with $\lambda_{FS}$ being the DM
free-streaming length). This eventually puts a strong upper bound on branching ratio of
inflaton decaying into RH neutrino and DM. To obtain the observed baryon asymmetry and at the same time
reproducing correct DM relic density, we have found that the DM mass should always be less than 320 MeV. Finally, we point out that non-thermal leptogenesis when combined with NeO opens up a new window in the RH mass scale where contrary to usual scenarios, the CP asymmetry parameter required for leptogenesis, decreases with the increase of RH mass scale. We call this window as a ``phantom window" (PW) and discuss importance of this PW in minimally fine tuned seesaw models. We do not present any explicit model and its possible  conformal UV completion. We restrict ourselves only into the detailed study of thermal vs. non-thermal production of lepton asymmetry/dark matter within the {\rm NeO}.\\

The rest of the paper is organised as follows. In Sec.\ref{s1}, we briefly discuss the framework of Neutrino Option. In Sec.\ref{s2} we discuss the inflaton decay to RH neutrinos and DM in a general context. Sec.\ref{s3} contains a discussion of inflaton decay within the Neutrino Option. We summarise our results in Sec.\ref{s4}.
\section{The neutrino option}\label{s1}
It is well known that in presence of the heavy RH neutrinos in the seesaw Lagrangian 
\bea
-\mathcal{L}^{\rm seesaw}= f_{\alpha i}\bar{\ell}_{L\alpha}\tilde{H} N_{Ri}
+\frac{1}{2}\bar{N}_{Ri}^C(M_R)_{ij} \delta _{ij}N_{Rj} 
+ {\rm h.c.}\,, \label{seesawlag}
\eea
where $l_{L\alpha}=\begin{pmatrix}\nu_{L\alpha} & e_{L\alpha}\end{pmatrix}^T$ is the SM lepton doublet of flavor $\alpha$, $\tilde{H}=i\sigma^2 H^*$ with $H= \begin{pmatrix}H^+&H^0
\end{pmatrix}^T $ being the Higgs doublet and $M_R={\rm diag}\hspace{.5mm} (M_1,M_2,M_3)$, $M_{1,2,3}>0$, the tree level Higgs potential 
\bea
V_0=-\frac{M_{H^0}^2}{2}H^\dagger H+\lambda_0 (H^\dagger H)^2
\eea
encounters large radiative correction ($\Delta M_H^2$ and $\Delta \lambda$) which monotonically increases with the mass scale of the heavy neutrinos. In particular, for the sake of naturalness\cite{n2}, the correction to Higgs mass not to exceed more than $\mathcal{O} (\rm TeV)$, one obtains an upper bound on the RH mass scale as $M\lesssim 10^7$ GeV\cite{n1}. The Neutrino Option\cite{Neo1,Neo2}, a similar idea, however assumes the classical potential to be nearly conformal at the UV (here at the overall scale of the heavy RH neutrinos $M$), i.e., 
\bea
M_{H^0}(\mu > M)\simeq 0, ~~\lambda_0 (\mu >M)\neq 0
\eea
with $\mu$ being the renormalization scale. The corrections are generated by the quantum effect of the RH neutrinos and hence break the invariance.  Thus the values of $M_{H}$ and $\lambda$ at Electroweak (EW) scale can be extrapolated with RGEs upto the mass scale $M$ to match the boundary conditions
\bea
M_{H}^2(\mu=M)\equiv \Delta M_H^2,~~\lambda(\mu=M)\equiv \lambda_0 + \Delta\lambda,
\eea
where in the limit of quasi-degenerate two heavy neutrinos, the radiative corrections $\Delta M_H^2$ and $\Delta \lambda$ computed with dimensional regularisation within  $\overline{\rm MS}$ renormalisation scheme \cite{Neo1} reads
\bea
\Delta M_H^2 (\mu = M)&=&\frac{1}{8\pi^2}{\rm Tr}\left[f M^2 f^\dagger\right],\label{thref}\\
\Delta \lambda (\mu = M) &=& -\frac{5}{32 \pi^2}\left( |f_1|^4+ |f_2|^4+2~ {\rm Re}(f_1.f_2^*)^2\right)-\frac{1}{16\pi^2}{\rm Im} (f_1.f_2^*)^2
\eea
with $f_i$ as the $i$th column of the Yukawa matrix $f$. Barring any fine tuning within the Yukawa entries (discussed later), the threshold correction to the Higgs mass can be re-cast as 
\bea
\Delta M_H^2 (\mu =M) = \frac{M_i^3}{4\pi^2 v^2}\overline{m},\label{higgs}
\eea
where $\overline{m}$ is a overall mass scale of the light neutrinos and is given by $\overline{m}=\frac{|f_i|^2v^2}{2M_i}$. For example, putting low energy neutrino data\cite{globalfit} and Higgs mass at EW scale, one obtains the RH mass scale $M\sim 10^7$ GeV. Due to insignificant running of the Higgs mass as well as light neutrino masses through SM RGEs (assuming the heavy states are decoupled and we are left with only seesaw-EFT\cite{rg1,rg2}), the relation in Eq.\ref{higgs} works well at low energies. Within this region of the parameter space the correction to $\lambda_0$ can also be neglected\cite{Neo2}. Eq.\ref{higgs} (with a more accurate version discussed in Sec.\ref{s3}) is the main constraint on the parameter space of leptogenesis in Neutrino Option. 
\section{The mechanism}\label{s2}

We first formulate the case of simultaneous non-thermal production of RH neutrinos and DM by considering a coupling \cite{port1,port2,port3,port4,port5}
\bea
\mathcal{L}^{N_i,DM}\sim y_\chi\overline{\chi}  \phi N_i, \hspace{2cm} {\rm with}~~i=1,2
\eea
where the field $\phi$ is the inflaton, $N_i$s are the two RH neutrinos and $\chi$ is a fermionic dark matter candidate. The case of pair production, i.e., $\phi \rightarrow N_i N_i$ will then be easy to understand and we shall mention it in relevant places alongside the present scenario.
We assume that after the end of inflation the energy density (${ n_\phi  m_\phi = \rho_r}=\pi^2 g_* T^4/30$) of inflaton field is transferred to the energy density of the radiation dominated universe. Thus the number densities of the RH neutrinos  and Dark Matter are given by
\bea
N_{N_1}\equiv N_{N_2}= B_{\chi}g_{\rm eff}\frac{\pi^4 { T_{\rm RH}}}{30  ~m_\phi},
N_{\rm DM}=B_{\chi}g_{\rm eff}\frac{\pi^4 {T_{\rm RH}}}{15 ~ m_\phi},\label{lepdark}
\eea
where $g_{\rm eff}=g_*/g_{N_i}$ with $g_*$ and $g_{N_i}$ as the total relativistic degrees of freedom and spin degrees of the RH neutrinos respectively.  In addition, we have assumed the branching rations of $\phi \rightarrow N_i\chi$ decays are the same (i.e., $B_\chi^{N_1}\equiv B_\chi^{N_2}\sim B_\chi$, so that $n_{N_i,\rm DM}\equiv B_\chi n_\phi$) and   $N_{N_{i,\rm DM}}$  are  number densities  normalised to the ultra-relativistic  equilibrium number densities ($ n_{i,\rm eq}^{\rm ur}=g_{N_i}T^3/\pi^2$) of $N_i$s. The Dark matter relic abundance\footnote{We are neglecting the annihilation cross sections of the processes like $N_iN_j\rightarrow \chi\chi$. Since our preferred value of $T_{\rm RH}$ is less than the RH neutrino masses, $\braket{\sigma v}_{N_iN_j\rightarrow \chi\chi}$ highly suppressed  and hence after reheating, it cannot generate appreciable freeze in density of $\chi$ to account for. Details are given in appendix.  } is given by\cite{density}
\bea
\hspace{-0.7cm}
\Omega_{\rm DM}h^2=\frac{M_{\rm DM} n_\gamma^0}{10.54 f({ T_{\rm RH}}, T_0){\rm GeVm^{-3}}} \left(\frac{N_{{\rm DM}}}{N_\gamma}\right)_{ T_{\rm RH}} \simeq 1.45 \times 10^6 \left(\frac{N_{{\rm DM}}}{N_\gamma}\right)_{ T_{\rm RH}} \left( \frac{M_{\rm DM}}{\rm GeV}\right),\label{relic}
\eea
where $n_\gamma^0\simeq 410.7\times 10^{6} \rm m^{-3}$ and $f( T_{\rm RH}, T_0)\simeq 27.3$ are the relic photon number density at the present time and photon dilution factor respectively. When combined with the latest Planck satellite  results~\cite{Aghanim:2018eyx} for  $\Omega_{\rm DM}h^2\simeq 0.12$, from Eq. \ref{relic} one finds
\bea
N_{N_{\rm DM}}\simeq 1.1\times 10^{-7} \left(\frac{\rm GeV}{ M_{\rm DM}} \right). \label{NDM}
\eea
Now from the second equation of Eq.\ref{lepdark} and Eq.\ref{NDM} one gets a generic expression for the reheating temperature $T_{\rm RH}$ as
\bea
T_{\rm RH}\simeq\frac{15\times 10^{-7}}{B_\chi g_{\rm eff}\pi^4} \left(\frac{\rm GeV}{ M_{\rm DM}} \right)  m_\phi. \label{Trh}
\eea
The $B-L$ asymmetry is given by\cite{bari}
\bea
N_{B-L}^{\rm lepto}=\sum_{i}^2\varepsilon_{i}\kappa_i\,, \label{bau0}
\eea
where  $\varepsilon_i$ is the unflavoured CP asymmetry parameter  corresponding to $i$th RH neutrino and $\kappa_i$ is the efficiency of the asymmetry production with an explicit expression given by\cite{bari,Samanta}
\bea
\kappa_i (z=M_{1}/T)=-\int_{z_{\rm T_{RH}}}^z \frac{dN_{N_i}}{dz^\prime}e^{-\sum_{i}\int_{z^\prime}^z W_{i}^{\rm ID+ S}(z^{\prime\prime})dz^{\prime\prime}}dz^\prime\, ,
\label{effi1}
\eea
where $W_{i}^{\rm ID+ S}$ includes the inverse decay and scattering processes that tend to washout the lepton asymmetry\cite{bari,nir}. Since we are assuming non-thermal production of the asymmetry, i.e.,  $T_{\rm RH}< M_{i}$, the washout effects ($W_{i}^{\rm ID+ S}$) are negligible. Thus the total lepton asymmetry is given by\cite{bari}
\bea
N_{B-L}=\sum_i N_{N_i}\biggr\rvert_{\rm RH} \varepsilon_i\equiv\frac{\pi^4 B_\chi g_{\rm eff}}{30}\sum_i \frac{\varepsilon_i T_{\rm RH}}{ m_\phi} \label{bau}
\eea
with $N_{B-L}^{\rm Obs}\simeq 6.1 \times 10^{-8}$\cite{planck}. Combining Eq.\ref{Trh} and Eq,\ref{bau} we obtain the master equation for the $B-L$ asymmetry as
\bea
N_{B-L}=N_{B-L}^{\rm Obs} (\varepsilon_1+\varepsilon_2) \left(\frac{\rm GeV}{ 1.22 M_{\rm DM}}\right)\equiv (\varepsilon_1+\varepsilon_2) \kappa_{\rm Dark}, \label{master}
\eea
where Eq.\ref{master} has been written in an analogous form as that of Eq.\ref{bau0}, though the physical meaning for the efficiency factors ($\kappa_i$ and $\kappa_{\rm Dark}$) are entirely different.\\
\begin{figure}
\includegraphics[scale=.39]{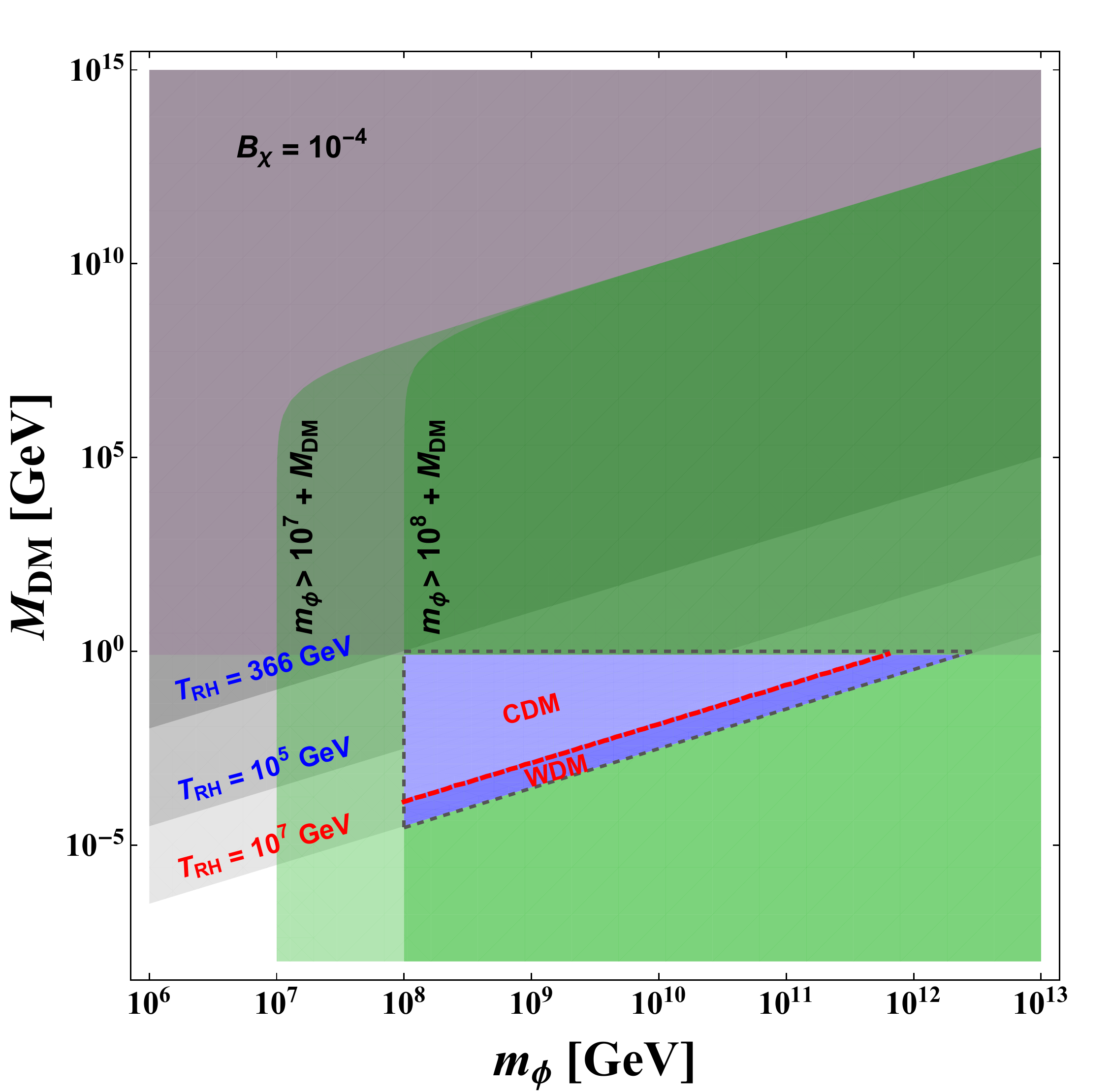}\includegraphics[scale=.39]{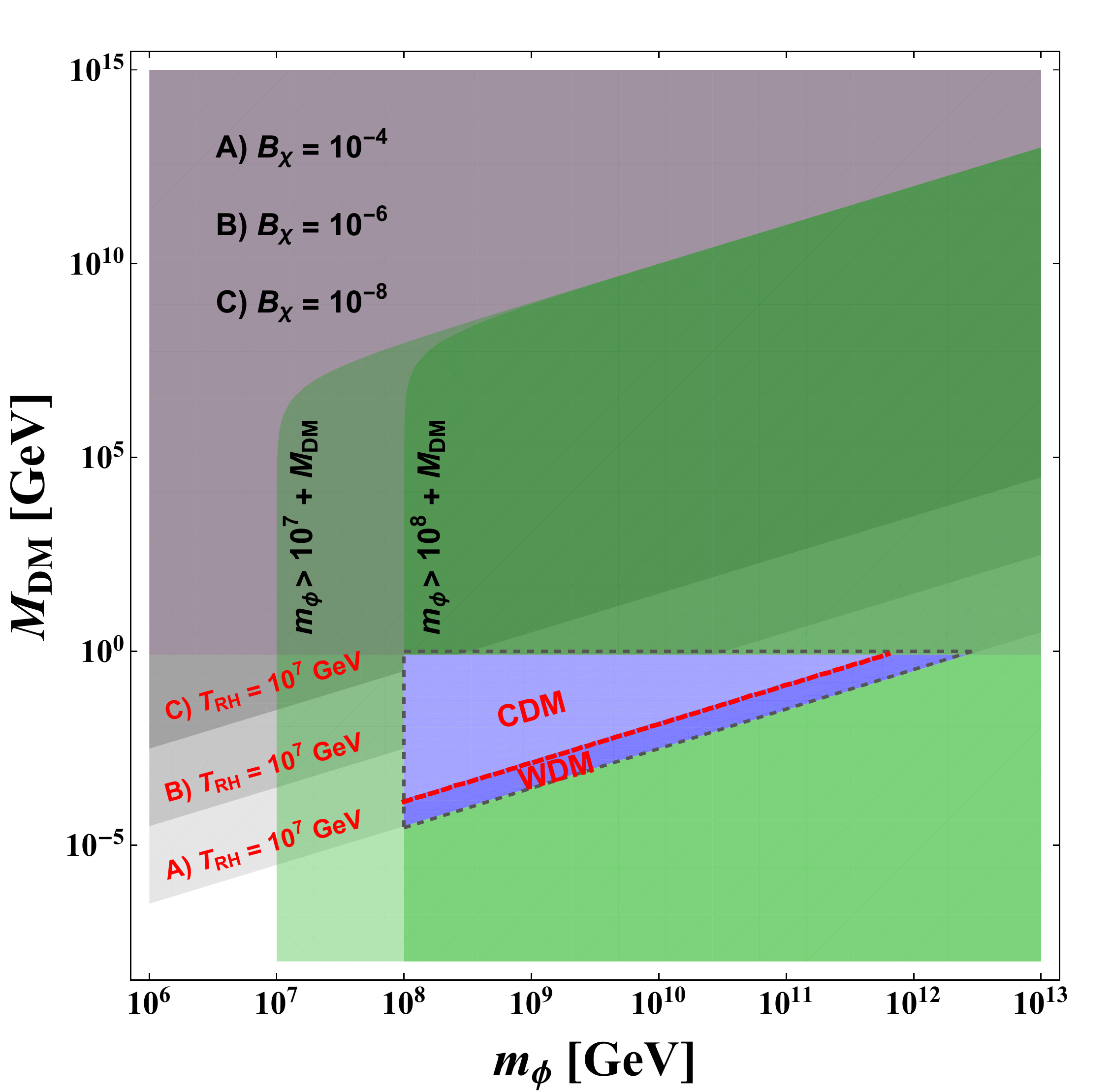}
\caption{Left: The green shades are the allowed region for the inflaton decay to be kinematically  allowed, i. e $m_\phi>M_{N_2}+M_{\rm DM}$. We have shown two green regions for two benchmark values of $M_{N_2}(=10^7~\rm GeV,10^8~\rm GeV)$. The pink shade from the top is the excluded region allowing $(\varepsilon_1+\varepsilon_2)^{\rm max }\approx 1$. The region bounded by the black dashed line is the maximum allowed region assuming $T_{\rm RH}^{\rm max}=10^7$ GeV. The two purple shades separated by a red dashed line represent the CDM ($\lambda_{\rm FS}\lesssim$ 10 kpc\cite{fs1,fs2}) and WDM region corresponding to  $T_{\rm RH}^{\rm max}$. Increasing gray gradients represent decrease in $T_{\rm RH}$ and hence truncation of the allowed parameter space. For $T_{\rm RH}$ = 366 GeV, the allowed parameter space closes which is consistent with Eq.\ref{ltrh}. It is evident that for $B_\chi=10^{-4}$ and for  $T_{\rm RH}=10^7$ GeV, correct relic density of DM is satisfied in the WDM region.  Right: Except the gray gradients, all the color codes are the same as in the left panel. Increasing  gray gradients represent decreasing  branching fraction and allowed area of the parameter space. Notice that once $B_\chi$ decreases, correct relic density of DM is satisfied in the CDM region.}\label{fig1}
\end{figure}

Non-thermal Dark matter can have large velocity ($v_D(t)$) at the matter radiation equality unless the $T_{\rm RH}$ is very high so that $v_D(t)$ is red-shifted\cite{nth3, devnth,Hisano:2000dz}. The large velocity might result into large co-moving free-streaming length (horizon) $\lambda_{\rm FS}$ which is constrained by the Ly-$\alpha$ cloud as $\lambda_{\rm FS}\lesssim 1$ Mpc\footnote{We use this value as a rough maximum allowed value of the free streaming length as considered in \cite{devnth}. However, in realistic case, where one needs to compute the momentum distribution function of the DM as well as the transfer function (beyond the scope of this study) relating the matter power spectra of WDM and CDM scenarios\cite{Boulebnane:2017fxw}, $\lambda_{FS}$ might be smaller. Since this section we set up the general scenario we use $\lambda_{FS}^{\rm max}\sim$ 1 Mpc. However, in section~\ref{s4} where we discuss the Neutrino Option we shall use a realistic value of $\lambda_{FS}$\cite{Boulebnane:2017fxw} to constrain the parameter space.}. An explicit expression for $\lambda_{\rm FS}$ is given by 
\bea
\lambda_{\rm FS}\equiv \int_{t_{\rm RH}}^{t_{\rm eq}} \frac{v_D(t)}{a(t)} dt,\label{fst}
\eea
where $a(t)\equiv a$ is the scale factor, $t_{\rm RH}$ and $t_{\rm eq}$ are  times at the reheating and matter radiation equality respectively. With the velocity $v_D(t)$ is given as
\bea
v_D(t)\approx \frac{\frac{m_\phi}{2}\frac{a_{T_{\rm RH}}}{a(t)}}{\sqrt{M_{\rm DM}^2+\frac{m_\phi^2 a_{T_{\rm RH}}^2}{4 a(t)^2}}}
\eea
and the Hubble expansion rate in the radiation domination\footnote{The actual expression
of $H$ in the radiation dominated era is $H = H_0 \sqrt{\Omega_R^0}\,\left(\dfrac{g_*(T)}{g_*(T_0)}\right)^{1/3}\,
a^{-2}$. Here we have neglected the ratio of energy degrees of freedom to get an analytical
expression of $\lambda_{\rm FS}$. This modifies $\lambda_{\rm FS}$ maximum by a factor of 3.}
$H \simeq H_0 \sqrt{\Omega_R^0} a^{-2}$ one simplifies Eq.\ref{fst} as
\bea
\lambda_{\rm FS}\simeq\frac{1}{H_0\sqrt{\Omega_R^0}}\int_{a_{T_{RH}}}^{a_{\rm eq}}\frac{da}{\sqrt{1+X^2 a^2}},\label{fst1}
\eea
where $X=\frac{2 M_{\rm DM}}{m_\phi a_{T_{\rm RH}}}$. Integrating Eq.\ref{fst1} the free streaming length $\lambda_{\rm FS}$ can be obtained as
\bea
\lambda_{\rm FS}\simeq \frac{\sqrt{\Omega_R^0}}{H_0 \Omega_M^0}P^{-1}\sinh^{-1}P,
\eea
where $P=a_{\rm eq}X$. Requiring $\lambda_{\rm FS}\lesssim$ 1 Mpc, with $H_0^{-1}\backsim 4.422\times10^3$ Mpc and $\Omega_M^0=0.3147$ from Planck 2018~\cite{Aghanim:2018eyx}, $t_{\rm eq}\backsim 2\times 10^{12} $ sec., and $t_{\rm RH}\backsim1/\Gamma_{\phi}^{\rm tot}\equiv (\pi^2g_*/30)^{-1/2}M_P/T_{\rm RH}^2$ one obtains a lower bound on the reheating temperature as
\bea
T_{\rm RH}\gtrsim 3.56\times 10^5 {\rm GeV}\left(g_*/200)\right) ^{-1/4} \left(\frac{\rm GeV}{ M_{\rm DM}}\right)\left(\frac{ m_\phi }{10^{12} \rm GeV}\right).
\eea
When combined with Eq.\ref{Trh}, the above translates in the upper bound on the branching ratio as
\bea
B_\chi \lesssim \frac{4.22}{\pi^4 g_{\rm eff}} \left(g_*/200)\right) ^{1/4}.
\eea
We can choose a representative value of $B_\chi=10^{-4}$, which satisfies the above upper limit. However, the  parameter space for the smaller  values for $B_\chi$ has been shown in the right panel of Fig.\ref{fig1}. One has also to take into account the constraints   on the  inflaton mass and the benchmark value (for which we want to discuss the model parameter space) of the reheating temperature ($T_{\rm RH} ^b$) as
\bea
 m_\phi^{\rm min} > M_{i}+ M_{\rm DM}, T_{\rm RH} ^b \lesssim M_{i}\,.\label{con1}
\eea
The second constraint in Eq.\ref{con1} when combined with Eq.\ref{Trh} translates into
\bea
 \left( \frac{ M_{\rm DM}}{\rm GeV}\right)\gtrsim\frac{3 \rm m_\phi}{10^4 \pi^4 T_{\rm RH}^b}. \label{con2}
\eea
Note that in Eq.\ref{con2},  we have used $B_\chi=10^{-4}$.
Finally, the maximum allowed value of the dark matter mass is given by Eq.\ref{master}, i.e.,
\bea
 \left( \frac{M_{\rm DM}}{\rm GeV}\right)\lesssim \frac{(\varepsilon_1 + \varepsilon_2)^{\rm max}}{1.22}\,.
 \label{con3}
\eea
Thus from Eq.\ref{con1}, Eq.\ref{con2} and Eq.\ref{con3} it is clear that in the $m_\phi$-$ M_{\rm DM}$ plane, the shape of the allowed parameter space is a triangle with area given by
\bea
\frac{\mathcal{A}}{\rm GeV^2}=\frac{1}{2}\left(\frac{(\varepsilon_1 + \varepsilon_2)}{1.22}-0.03 \frac{ m_\phi^{\rm min}}{10^4T_{\rm RH}^b}\right)\left(3.2\times 10^6  \frac{(\varepsilon_1 + \varepsilon_2)}{1.22} T_{\rm RH}^b- m_\phi ^{\rm min}\right)\,.
\label{area}
\eea 
From Eq.\ref{area}, one obtains a lower bound on the $T_{\rm RH}^b$ as 
\bea
T_{\rm RH}^b\gtrsim \frac{0.0366 ~\rm m_\phi^{\rm min}}{10^4 (\varepsilon_1 + \varepsilon_2)}.\label{ltrh}
\eea
For example, in an extreme fine tuned condition of $(\varepsilon_1 + \varepsilon_2)^{\rm max}\approx 1$ and a benchmark value $\rm m_\phi^{\rm min}\sim 10^8$ GeV one obtains the lower bound on the reheating temperature in this scenario\footnote{The benchmark value $B_\chi=10^{-4}$ can be slightly relaxed by demanding  $T_{\rm RH}\simeq T_{\rm Sph}$ with $T_{\rm Sph}$ being the Sphaleron freeze out  temperature. In that case, one obtains $B_\chi=\frac{0.03 m_\phi}{T_{\rm Sph} 10^8 (\varepsilon_1+\varepsilon_2)^{\rm max}}$.  Thus for our preferred set of values, $B_\chi$ can be relaxed to $B_\chi \sim 2\times 10^{-4}$, where we use $T_{\rm Sph}=132$ GeV. } as $T_{\rm RH}^{\min}=366$ GeV. In Fig.\ref{fig1} we show the maximum allowed parameter space for a minimum bench mark value of inflaton mass $m_\phi^{\rm min}=10^8$ GeV. The red line of  which separates the regions of  Cold Dark Matter (CDM) and the Warm Dark Matter (WDM), corresponds the free streaming length $\lambda_{\rm FS}=10$ kpc, a value taken from Ref.\cite{fs1,fs2} which was also used in Ref.\cite{devnth}.  We would like to point out that while deriving  lower bound on the reheat temperature with Eq.\ref{Trh}, one has to be a bit more careful. This is since, in the left panel of Fig.\ref{fig1} we assume whatever be the choices of $T_{\rm RH}$, the CP asymmetry parameter $\varepsilon_1+\varepsilon_2$ remains constant. However, since the choice of $T_{\rm RH}$   depends on the RH masses and so is $\varepsilon_1+\varepsilon_2$, one has to consider these two effect simultaneously. We properly consider this while extracting the parameter space for leptogenesis in Neutrino Option. Note that in the case of pair production $\phi\rightarrow N_i N_i$ for which RHS of Eq.\ref{bau} will be enhanced by a factor 2, there is no such constraints on the branching ratio $B_\chi$\cite{nth6,shafi}. Thus one expects larger parameter space in this case as we shall show in the next section. Having set up all the necessary prerequisites we now move to the discussion of non-thermal lepton asymmetry and DM production in the context of Neutrino Option.

\section{The scenario of $\phi\rightarrow N_i \chi, N_i N_i$ decays within the neutrino option }\label{s3}
Starting  from the neutrino part of the  seesaw Lagrangian in Eq.\ref{seesawlag}
\bea
-\mathcal{L}_{mass}^{\nu,N}= \bar{\nu}_{L\alpha}(m_D)_{i\alpha}N_{Ri}
+\frac{1}{2}\bar{N}_{Ri}^C(M_R)_{ij} \delta _{ij}N_{Rj} 
+ {\rm h.c.}\,, 
\eea
the effective light neutrino mass matrix can be obtained with the seesaw mechanism\cite{sw1} as
\bea
M_\nu = -m_DM_R^{-1}m_D^T\,. \label{seesaweq}
\eea
The mass matrix in Eq.\ref{seesaweq} can be put into diagonal from by a unitary matrix $U$ as
\bea
U^\dagger m_D M_R^{-1}m_D^T U^*=D_m\label{see2}
\eea
where $D_m=-~{\rm diag}~(m_1,m_2,m_3)$ with $m_{1,2,3}$ being the physical light neutrino masses. We work in a basis where the charged lepton mass matrix $m_\ell$ and the RH neutrino mass matrix $M_R$ are diagonal. Thus, the neutrino mixing matrix $U$ can be written as 
\bea
U=P_\phi U_{PMNS}\equiv 
P_\phi \begin{pmatrix}
c_{1 2}c_{1 3} & e^{i\frac{\alpha}{2}} s_{1 2}c_{1 3} & s_{1 3}e^{-i(\delta - \frac{\beta}{2})}\\
-s_{1 2}c_{2 3}-c_{1 2}s_{2 3}s_{1 3} e^{i\delta }& e^{i\frac{\alpha}{2}} (c_{1 2}c_{2 3}-s_{1 2}s_{1 3} s_{2 3} e^{i\delta}) & c_{1 3}s_{2 3}e^{i\frac{\beta}{2}} \\
s_{1 2}s_{2 3}-c_{1 2}s_{1 3}c_{2 3}e^{i\delta} & e^{i\frac{\alpha}{2}} (-c_{1 2}s_{2 3}-s_{1 2}s_{1 3}c_{2 3}e^{i\delta}) & c_{1 3}c_{2 3}e^{i\frac{\beta}{2}}
\end{pmatrix}\,,\nonumber\\
\label{eu}
\eea
where $P_\phi={\rm diag}~(e^{i\phi_1},~e^{i\phi_2}~e^{i\phi_3})$ is an unphysical diagonal phase matrix and $c_{ij}\equiv\cos\theta_{ij}$, $s_{ij}\equiv\sin\theta_{ij}$ with the mixing angles $\theta_{ij}=[0,\pi/2]$. Low energy CP violation enters in Eq.\,\ref{eu} through the Dirac phase $\delta$ and the Majorana phases $\alpha$ and $\beta$.  As an aside, it is convenient to parametrise (which can be straightforwardly derived from Eq.\ref{see2}) the Dirac mass matrix as
\bea
m_D=U\sqrt{D_m}\Omega\sqrt{M_R},\label{orth}
\eea
where $\Omega$ is a $3\times 3$ complex orthogonal matrix that contains high energy CP phases. The parametrisation of the Dirac matrix, known as the Cassas-Ibarra parametrisation\cite{CI} is an important and useful parametrisation particularly in the studies of leptogenesis, since depending upon the structure of the orthogonal matrix it becomes easier to understand whether leptogenesis is driven by the low energy or high energy CP phases. Before going into the detail discussion of leptogenesis in the scenario under consideration, in Table \ref{oscx}, let's present the latest fact file for the light neutrinos.
\begin{table}[H]
\begin{center}
\caption{Input values used in the analysis (inclusive of SK data)\cite{globalfit}} \label{t1}
\vspace{2mm}\label{oscx}
 \begin{tabular}{|c|c|c|c|c|c|}
\hline
\hline
${\rm Parameter}$&$\theta_{12}$&$\theta_{23}$ &$\theta_{13}$ &$ \Delta
m_{21}^2$&$|\Delta m_{31}^2|$\\
&$\rm degrees$&$\rm degrees$ &$\rm degrees$ &$ 10^{-5}\rm
(eV)^2$&$10^{-3} \rm (eV^2)$\\
\hline
$3\sigma\hspace{1mm}{\rm
ranges\hspace{1mm}(NO)\hspace{1mm}}$&$31.61-36.27$&$41.1-51.3$&$8.22-8.98$&
$6.79-8.01$&$2.44-2.62$\\
\hline
$3\sigma\hspace{1mm}{\rm
ranges\hspace{1mm}(IO)\hspace{1mm}}$&$31.61-36.27$&$41.1-51.3$&$8.26-9.02$&
$6.79-8.01$&$2.42-2.60$\\
\hline
${\rm Best\hspace{1mm}{\rm fit\hspace{1mm}}values\hspace{1mm}(NO)}$ &
$33.82$ & $48.6$ &  $8.60$ &$7.39$ & $2.53$\\
\hline
${\rm Best\hspace{1mm}{\rm
fit\hspace{1mm}}values\hspace{1mm}(IO)}$&$33.22$&$48.8$&$8.64$&$7.39$&$2.51$\\
\hline
\end{tabular}
\end{center}
\end{table}
Compared to the previously released data\cite{gfpre}, present best-fit value ($\sim 221^o$) for the Dirac CP violating phase ($\delta$) exhibits a shift towards its CP conserving value for the Normal mass Ordering (NO), however for the Inverted mass Ordering (IO), best-fit of $\delta$ is still close to its maximal value ($\sim 282^o$). The Majorana phases remain unconstrained and there is preference of a NO over an IO.\\

Before the Electroweak Symmetry Breaking (ESB), the RH neutrinos decay to lepton doublets and Higgs (cf. Eq.\ref{seesawlag}). In general, the produced lepton doublets $\ket{\ell_i}$  can be written as a coherent superposition of the corresponding flavour states $\ket{\ell_{\alpha}}$ as,
\bea
\ket{\ell_i}&=&\mathcal{A}_{i\alpha} \ket{\ell_\alpha} \hspace{1cm} (i=1,2; \alpha=e,\mu,\tau)\\
\ket{\bar{\ell}_i}&=&\bar{\mathcal{A}}_{i\alpha} \ket{\bar{\ell}_\alpha} \hspace{1cm} (i=1,2; \alpha=e,\mu,\tau)\,,
\eea
where the tree level amplitudes are given by
 \bea
 \mathcal{A}_{i\alpha}^0 =\frac{({m_D})_{i\alpha}}{\sqrt{(m_D^\dagger m_D)_{ii}}}\hspace{1cm}{\rm and}\hspace{1cm}\bar{\mathcal{A}}_{i\alpha}^0 =\frac{({m^*_D})_{i\alpha}}{\sqrt{(m_D^\dagger m_D)_{ii}}}.\label{states}
 \eea
 However, the mass regime we are working in ($M_i \lesssim \rm 10~PeV$), the charged lepton interactions are strong enough\cite{fl1,fl2,fl3,fl4} to completely break the coherence of  $\ket{\ell_i}$ states and thereby resolve each of the flavours. Thus one has to track the asymmetry in individual flavours.
 The asymmetry in the flavour $\alpha$ is given by
\bea
N_{\Delta_{\alpha}}=\sum_{i}^2\varepsilon_{i\alpha}\kappa_{i\alpha}
\label{fep}
\eea
with the efficiency factor
\bea
\kappa_{i\alpha} (z)=-\int_{z_{\rm in}}^z \frac{dN_{N_i}}{dz^\prime}e^{-\sum_{j}\int_{z^\prime}^zP_{j\alpha}^0 W_{j}^{\rm ID}(z^{\prime\prime})dz^{\prime\prime}}dz^\prime\,,\label{effi}
\eea
where $P^0_{i\alpha}=|\mathcal{A}^0_{i\alpha}|^2$. With this definition, the total $N_{B-L}$ asymmetry is given by 
\bea
N_{B-L}=\sum_{\alpha}N_{\Delta_{\alpha}}\,. 
\eea
As mentioned earlier, since in our case the production is non-thermal, the wash-out of the asymmetry is negligible and thus the efficiency factor $\kappa$ is basically independent of the flavour index `$\alpha$'. Therefore the total Lepton asymmetry is 
\bea
N_{B-L}=\sum_{\alpha}N_{\Delta_{\alpha}}\equiv \sum_{i} \kappa_{i}\sum_{\alpha}\varepsilon_{i\alpha}\simeq \sum_{i}N_{N_i}\biggr\rvert_{\rm RH} \varepsilon_{i}.\label{tnbl}
\eea

The flavoured CP asymmetry parameter is given by\cite{pilaf}
\bea
\varepsilon_{i\alpha}
&=&-\frac{1}{4\pi v^2 h_{ii}}\sum_{j\ne i}\left[ {\rm Im}\{h_{ij}
({m_D^\dagger})_{i\alpha} (m_D)_{\alpha j}\} g({x_{ij})}
-\frac{(1-x_{ij})
{\rm Im}\{{h}_{ji}({m_D^\dagger})_{i\alpha} (m_D)_{\alpha j}\}}
{(1-x_{ij})^2+{{h}_{jj}^2}{(16 \pi^2 v^4)}^{-1}}\right],\nonumber \\
\label{ncp}
\eea 
where $h_{ij}=(m_D^\dagger m_D)_{ij}$, $x_{ij}=M_j^2/M_i^2$ and $g(x_{ij})$ is given by
\bea
g(x_{ij})=\left[\sqrt{x_{ij}}[1-(1+x_{ij})~{\rm ln}\left(\frac{1+x_{ij}}{x_{ij}}\right)]+\frac{\sqrt{x_{ij}}(1-x_{ij})}
{(1-x_{ij})^2+{{h}_{jj}^2}{(16 \pi^2 v^4)}^{-1}}\right]. \label{gxij}
\eea
Since $h_{ij}$ is a hermitian matrix, when summed over $\alpha$, the second term in Eq.\ref{ncp} vanishes. Using the orthogonal parametrisation for $m_D$ given in Eq.\ref{orth}, the total CP asymmetry parameter (which is relevant in our case, cf. Eq.\ref{tnbl}) can be written as 
\bea
\varepsilon_i &= &-\frac{1}{4\pi v^2}\sum_{\alpha}\frac{{\rm Im} [M_j\sum_{k k^\prime}\sqrt{m_k m_{k^\prime}}m_k \Omega^*_{ki}\Omega^*_{k^\prime i} U^\dagger_{k^\prime \alpha}U_{\alpha k}]g(x_{ij})}{\sum_{k^{\prime \prime}}m_{k^{\prime \prime}}|\Omega_{k^{\prime\prime} i}|^2}\,, \nonumber\\
&=& -\frac{1}{4\pi v^2}\frac{ M_jg(x_{ij})\sum_{k}m_k^2{\rm Im} [ \Omega^*_{ki} \Omega^*_{ki}  ]}{\sum_{k^{\prime \prime}}m_{k^{\prime \prime}}|\Omega_{k^{\prime\prime} i}|^2} ~~~~{\rm with}~~i,j (i\neq j)=1,2.\label{cporth} 
\eea
Note that the total CP asymmetry parameter given in Eq.\ref{cporth} is independent of light neutrino mixing angles and low energy CP violating phases. Thus this scenario is insensitive in general  to neutrino experiments. In addition, the models which predict a real or purely imaginary orthogonal matrix, see e.g., Refs.\cite{reo1,reo2,reo3,reo3a,reo4,reo5,reo6}, can not generate baryon asymmetry via leptogenesis. Thus non-thermal leptogenesis scenario is not compatible with models of those kind.
For simplicity, we shall work in a two RH neutrino scenario assuming the third one is heavier enough to be decoupled from the seesaw formula. In that case, the orthogonal matrices  for NO ($m_1=0$) and IO ($m_3=0$) are given by
\bea
\Omega^{\rm NO}=\begin{pmatrix}
0&0&1\\ \cos\theta &\sin\theta &0\\-\sin\theta & \cos\theta & 0
\end{pmatrix},\hspace{1cm} \Omega^{\rm IO}=\begin{pmatrix}
 \cos\theta &\sin\theta &0\\-\sin\theta & \cos\theta & 0\\0&0&1
\end{pmatrix}, \label{2orth}
\eea
where $\theta=x-iy$ is a complex angle with $x$ and $y$ being real parameters. Using Eq.\ref{2orth} and maximising Eq.\ref{cporth} with respect to $x$ we get\footnote{Since our aim is to extract maximum parameter space, we maximize the CP asymmetry parameter by putting $x=\pi/4$.}
\bea
\varepsilon_1^{\rm NO}=-\frac{M_2 g(x_{12})}{4\pi v^2} (m_3-m_2) \tanh 2y,\label{en1} \\
\varepsilon_2^{\rm NO}=\frac{M_1 g(x_{21})}{4\pi v^2} (m_3-m_2) \tanh 2y 
\eea
and
\bea
\varepsilon_1^{\rm IO}=-\frac{M_2 g(x_{12})}{4\pi v^2} (m_2-m_1) \tanh 2y, \\
\varepsilon_2^{\rm IO}=\frac{M_1 g(x_{21})}{4\pi v^2} (m_2-m_1) \tanh 2y.\label{ei2}
\eea

From Eq. \ref{thref}, the threshold correction to the Higgs mass $\Delta M_H^2$ can be written as 
\bea
\Delta M_H^2=\frac{1}{8\pi^2 v^2}\sum_{i} M_i^3 \sum_k m_k |\Omega_{ki}|^2=\frac{m^*}{8\pi^2 v^2}\sum_{i} M_i^3 K_i\label{thre0}\,
\eea
where, the decay parameter for the $i^{\rm th}$ RH neutrino
is defined as $K_i = \frac{\sum_k m_k |\Omega_{ki}|^2}{m^*}$ and $m^*$ is the equilibrium
neutrino mass given by $m^*\sim 10^{-3}$ eV\cite{nir}. The above expression 
in the quasi degenerate limit of the RH neutrino masses ($M_1\simeq M_2 \simeq M$) simplifies to
\bea
\Delta {M_H^2}^{\rm NO}\simeq \frac{M^3}{8\pi^2 v^2} (m_3+m_2) \cosh 2y,~
\Delta {M_H^2}^{\rm IO}\simeq \frac{M^3}{8\pi^2 v^2} (m_2+m_1) \cosh 2y\,.
\label{thre}
\eea
Thus using Eq.\ref{thre} and Eq.\ref{en1}-Eq.\ref{ei2},
the total CP asymmetry parameters can be simplified as 
\bea
(\varepsilon_1+\varepsilon_2)^{\rm NO}=-\frac{m_3-m_2}{4\pi v^2}[g(x_{12})M_2-g(x_{21})M_1]F^{\rm NO}(M_H,M,\sum_i m_i),\\
(\varepsilon_1+\varepsilon_2)^{\rm IO}=-\frac{m_2-m_1}{4\pi v^2}[g(x_{12})M_2-g(x_{21})M_1]F^{\rm IO}(M_H,M,\sum_i m_i),\label{noeps}
\eea
where 
\bea
F^{\rm NO}(M_H,M,\sum_i m_i)=\sqrt{1-\left[\frac{M^3(m_3+m_2)}{\Delta M_H^2 8\pi^2 v^2}\right]^2},\\F^{\rm IO}(M_H,M,\sum_i m_i)=\sqrt{1-\left[\frac{M^3(m_2+m_1)}{\Delta M_H^2 8\pi^2 v^2}\right]^2}.
\eea
\begin{figure}
\includegraphics[scale=.7]{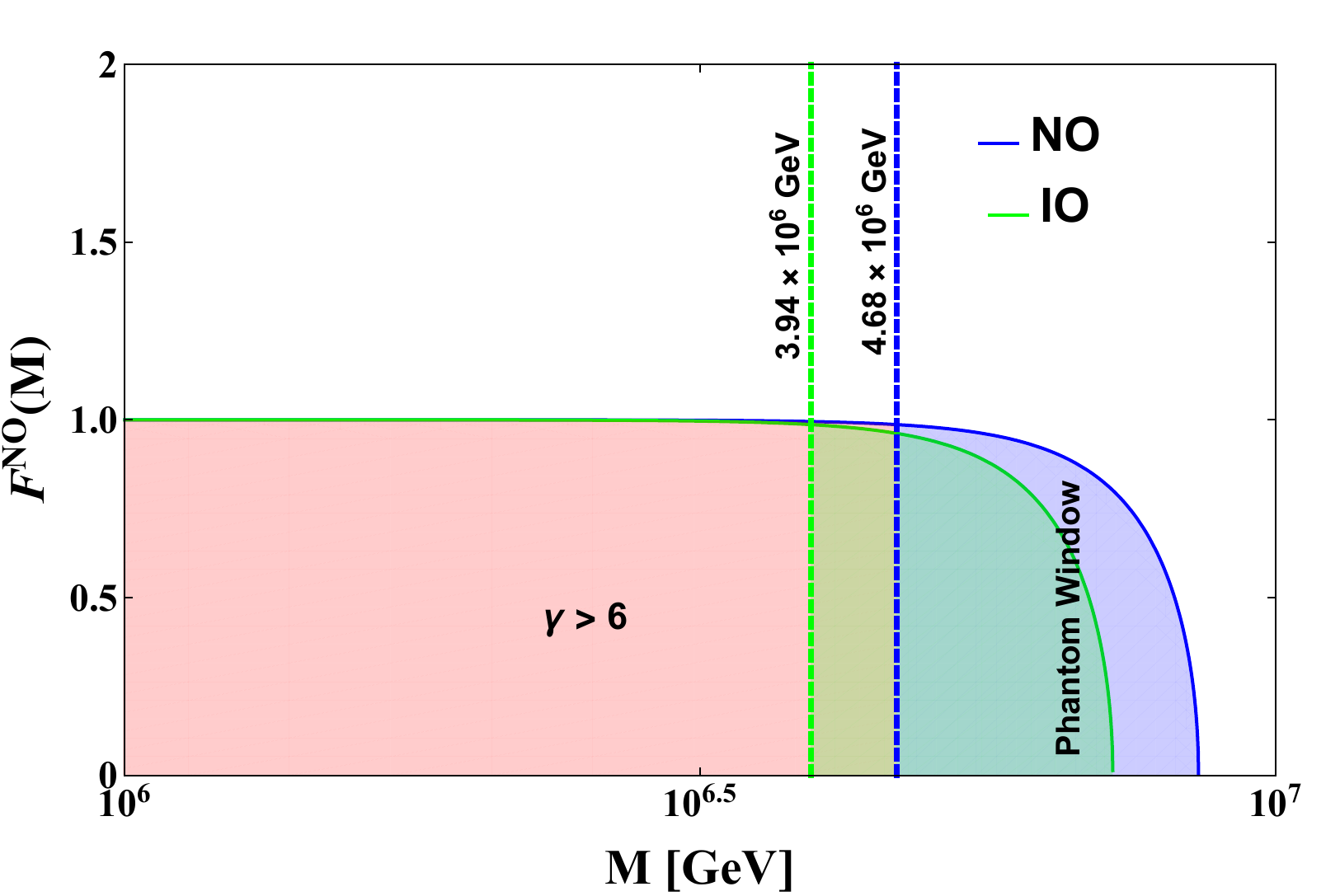}
\caption{Variation of the phantom function with the RH mass scale $M$. The green (for inverted ordering) and blue (for normal ordering) lines represent  starting point of the phantom window which causes a decrease of the CP asymmetry parameter as the RH mass scale increases. The pink region belongs to boosted seesaw systems in which the phantom function saturates to 1. }\label{phn}
\end{figure}
The function $F^{\rm (N,I)O}$, we call it a phantom function, renders a window to the RH mass scale (cf. Fig.\ref{phn}) in which it drives  the CP asymmetry parameters to decrease with the increase of the RH mass scale. Furthermore, $\varepsilon_i$ vanishes at $M_{\rm max}$ which can be calculated for $y=0$ in Eq.\ref{thre} for the respective ordering of light neutrino masses as $M^{\rm NO}_{\rm max}\simeq 8.57\times 10^6$ GeV and $M^{\rm IO}_{\rm max}\simeq 7.21\times 10^6$ GeV. Thus as the RH mass scale increases, in this window, one needs the RH neutrinos to be more quasi-degenerate to resonantly amplify the CP asymmetry parameter. This is a completely new behaviour of the CP asymmetry parameter introduced by the Neutrino Option. To stress more on the significance  of the phantom window (PW) in seesaw models, let's have a brief look at the properties of the orthogonal matrix $\Omega$. First of all, the quantity 
\bea
\gamma_i=\sum_j |\Omega^2_{ij}|\geq 1\label{boost}
\eea
accounts for the fractional contribution of the heavy $M_j$ states to a particular light neutrino $m_i$, thus can be treated as a measure of fine-tuning in the seesaw formula\cite{haar}. In addition, since $\Omega$ belongs to $SO(3,\mathbb{C})$, it is isomorphic to the Lorentz group. Thus $\Omega$ can be factorized as 
\bea
\Omega = \Omega^{\rm rotation}\Omega^{\rm Boost}.
\eea
Using Eq.\ref{orth} and Eq.\ref{states} it is trivial to derive a transformation relation between the states produced by the RH neutrinos ($\ket{\ell_j}$) and the light neutrinos states ($\ket{\tilde{\ell}_i}$) as
\bea
\ket{\ell_j} =B_{ji}\ket{\tilde{\ell}_i},
\eea
where the bridging matrix $B_{ij}$, first introduced in Ref.\cite{haar} relates the heavy and the light states with a non-orthonormal transformation (in general) and is related to the orthogonal matrix as 
\bea
B_{ji}=\frac{\sqrt{m_i}\Omega_{ji}}{\sqrt{m_k|\Omega_{kj}|^2}}.
\eea
For a simple choice of the orthogonal matrix $\Omega$ which does not corresponds to any fine-tuning (a particular heavy neutrino contributes to a particular light neutrino\cite{form}), e.g.,
\bea
\Omega^{\rm NO}=\begin{pmatrix}
0&0&1\\ 1 &0 &0\\0 & 1 & 0
\end{pmatrix},
\eea
the heavy and the light states coincide as shown in Fig.\ref{bases}.
\begin{figure}
\includegraphics[scale=.5]{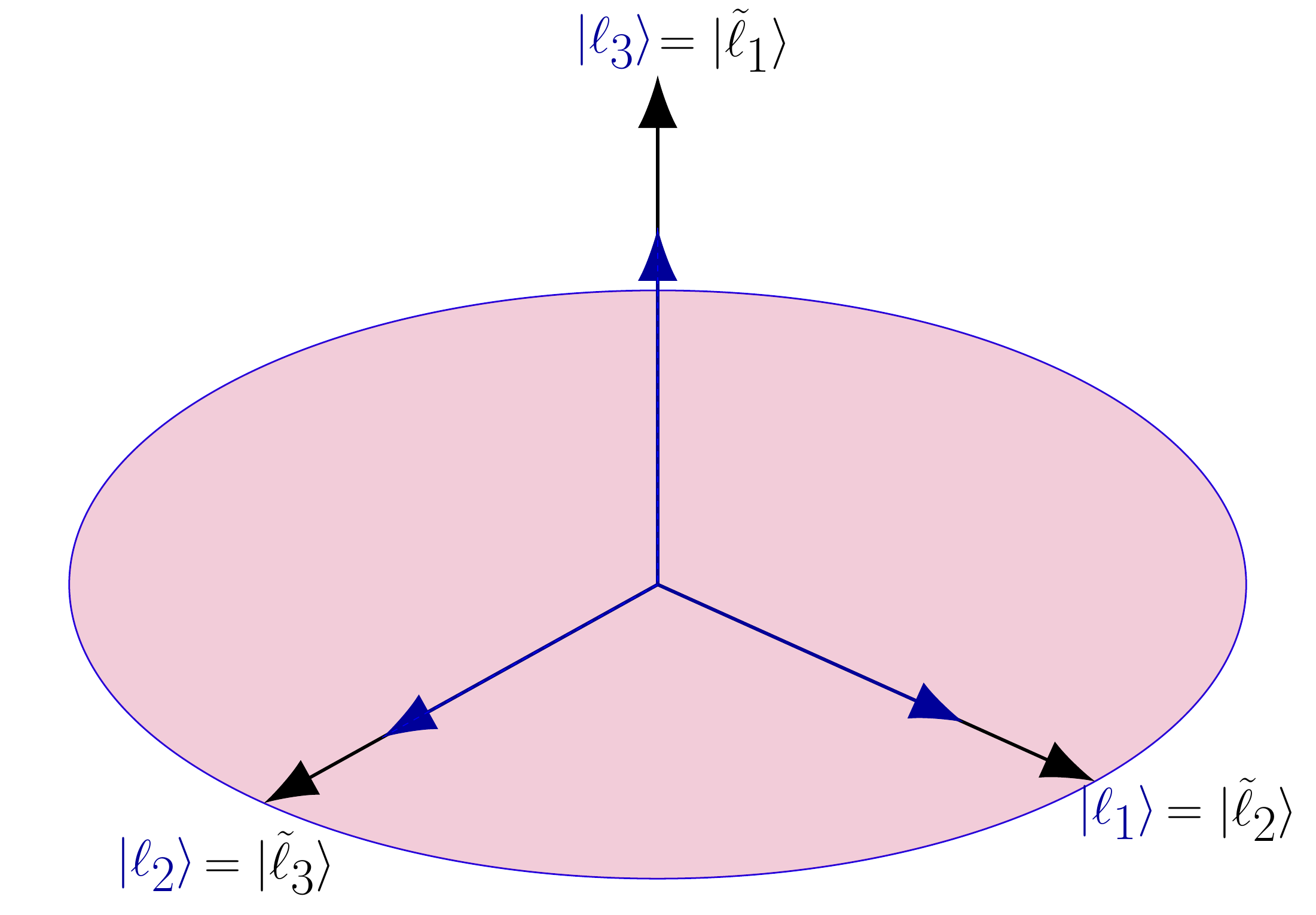}
\caption{Illustration of seesaw models with no fine tuning leading to vanishing lepton asymmetry. The state vectors in blue are heavy states produced by the RH neutrinos while the states in black correspond to the light neutrinos.} \label{bases}
\end{figure}
However this is not true in general. For the  orthogonal matrix (cf.  Eq.\ref{2orth}) which can be factorised as
\bea
\Omega^{\rm NO}=\begin{pmatrix}
0&0&1\\ \cos x &\sin x &0\\-\sin x & \cos x & 0
\end{pmatrix} \begin{pmatrix}
 \cosh y &-i \sinh y &0\\i \sinh y & \cosh y & 0\\0&0&1
\end{pmatrix},
\eea
the orthonormality in the heavy states does not hold unless one assumes $x,y=0$. Specifically, due to the presence of the boost matrix, the heavy states are in general strongly non-orthonormal. Using Eq.\ref{boost} the fine-tuning parameters can be  can be calculated as 
\bea
\gamma_2=\gamma_3=\cosh 2y.
\eea
This explicitly shows how the boosted seesaw systems with strongly non-orthonormal heavy states may involve large amount of fine-tuning and the philosophy, that the separation between two mass scales $M$ and $M_H$ (cf. Eq.\ref{thre}) may be related to another experimentally observed light mass scale $m_i$, gets spoiled. From now on we call the $`\gamma_i$' parameters as the boost parameters and the seesaw systems which correspond to $\gamma_i\approx 1$ as the minimally fine-tuned seesaw models or unboosted seesaw systems. In Fig.\ref{phn}, we show the variation of the phantom function $F^{\rm (N,I)O}$ with the RH mass scale $M$. For the best fit values of the light neutrino masses given in Table \ref{t1}, we find
\bea
M^{\rm NO}_F= 4.68\times 10^6 {\rm GeV},~~M^{\rm IO}_F= 3.94\times 10^6 {\rm GeV},\label{mmax}
\eea
below which the function $F^{\rm (N,I)O}$ saturates approximately to 1. Thus $M^{\rm NO}_F$ and $M^{\rm IO}_F$ can be regarded as the starting point of PW. Using Eq.\ref{thre}, one finds, $M\lesssim M^{(\rm N,I)O}_F$ corresponds to $\gamma_i \gtrsim 6$ which shows even the starting point of the PW corresponds to large fine-tuning in the seesaw formula. 
\begin{figure}
\includegraphics[scale=.35]{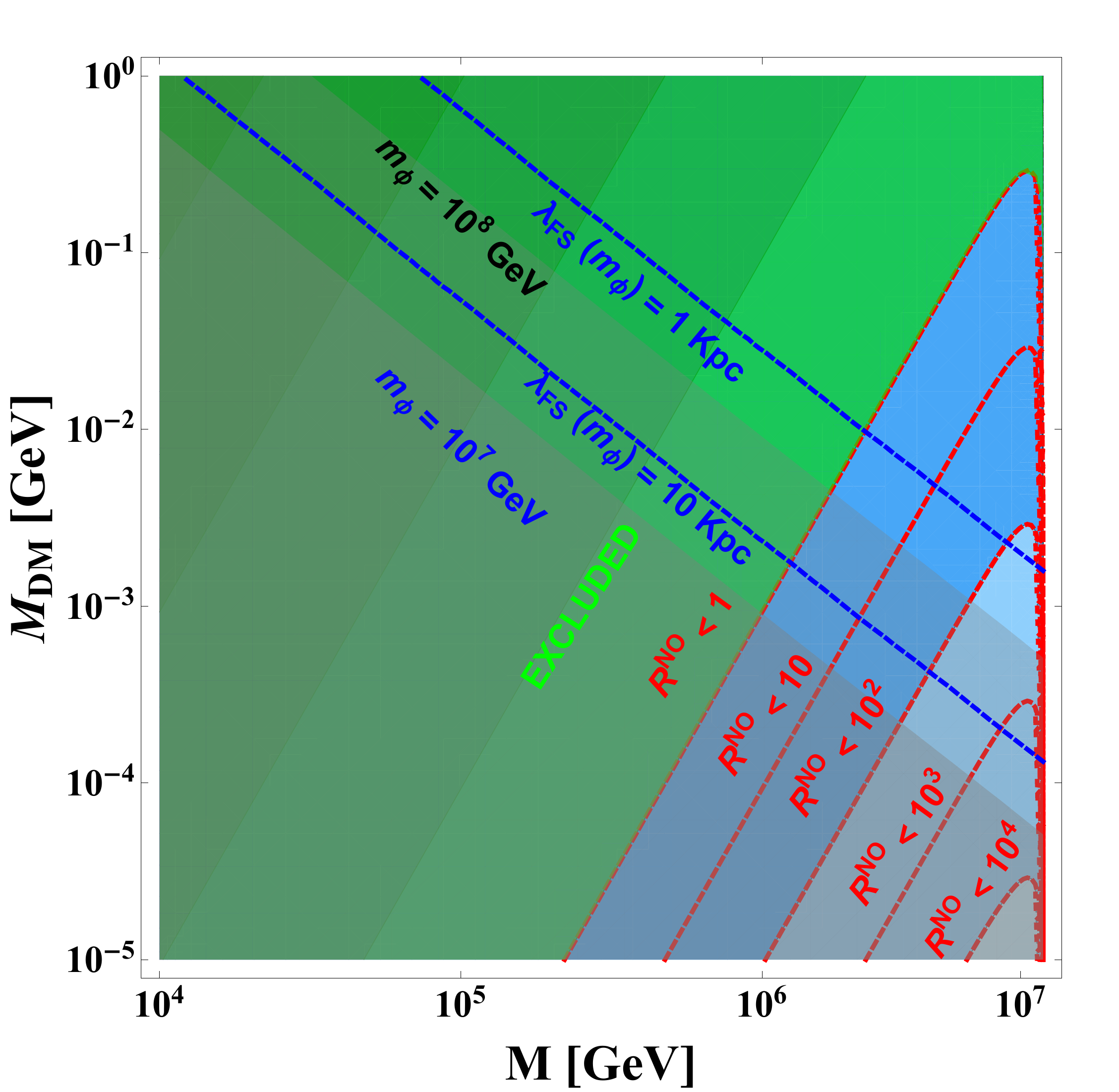}\includegraphics[scale=.35]{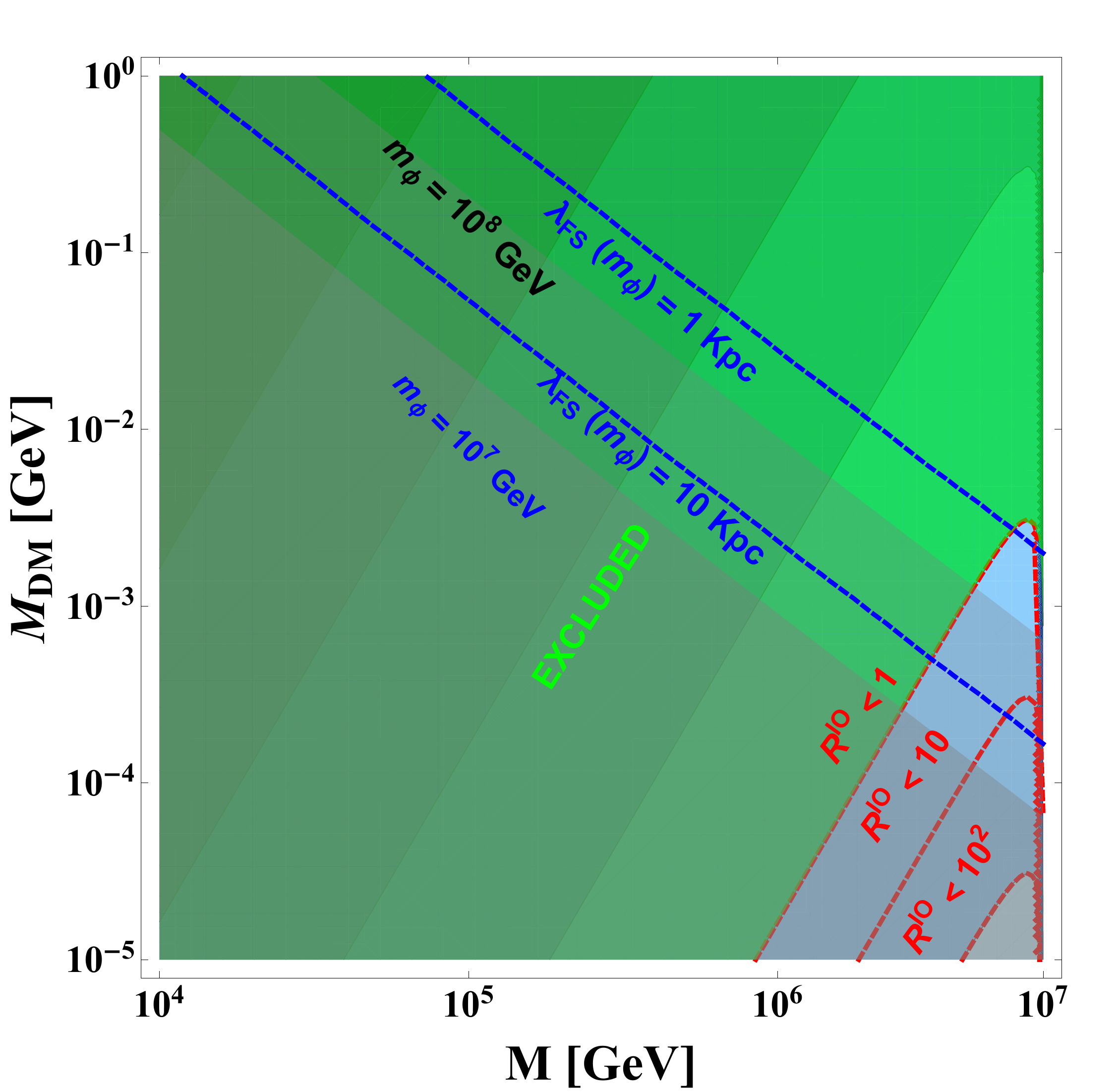}\\
\includegraphics[scale=.35]{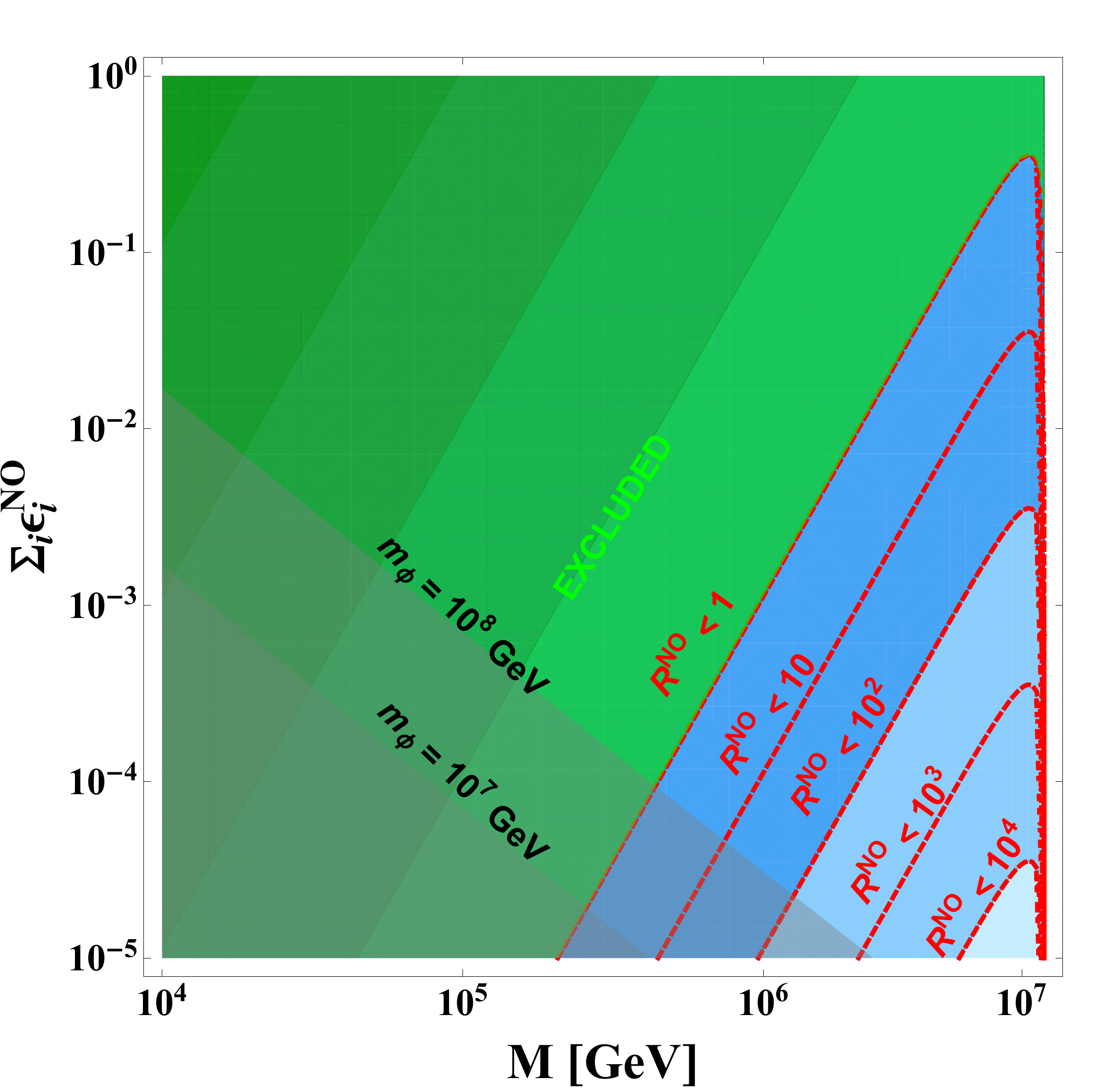}\includegraphics[scale=.35]{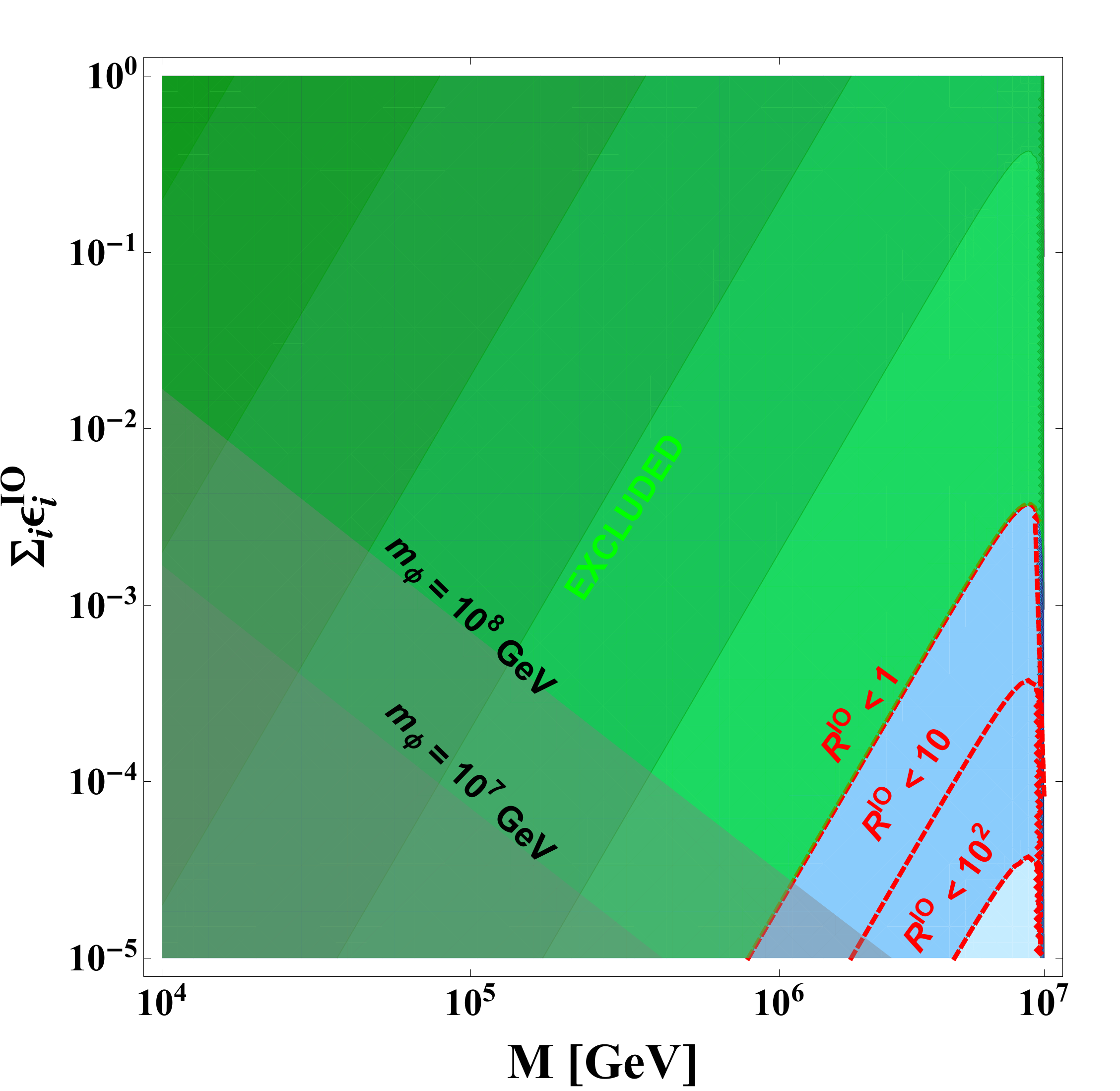}
\caption{In the $M-M_{\rm DM}$ plane, allowed values of the $R$ parameter which measures the departure from Pilaftsis-Underwood resonance  for normal ordering (left) for inverted ordering (right). The green region is excluded as it corresponds to a value $R<1$.}\label{pspace}
\end{figure}

Now we turn to the detail analysis of the model parameter space which satisfies correct DM relic density as well as observed baryon asymmetry (for the $\phi\rightarrow N_i \chi$ decay). First of all, the regulator term in Eq.\ref{ncp} can be written as 
\bea
\tilde{\Gamma}_2= h_{22}/4\pi v^2\simeq 1/ 4\pi v^2 \sum_{k}m_k|\Omega_{k2}|^2 M_2 \equiv\frac{m^*}{4\pi v^2} K_2 M_2\simeq \tilde{\Gamma}_1.
 \eea
With $\delta_{\rm lep}=(M_2-M_1)/M_1$, we  define  ratios
\bea
R^{\rm NO}=\frac{\delta_{\rm lep}}{\tilde{\Gamma}_2^{\rm NO}}, ~~R^{\rm IO}=\frac{\delta_{\rm lep}}{\tilde{\Gamma}_2^{\rm IO}}
\eea
which measure the departure from a pure Pilaftsis-Underwood resonance ($R^{\rm (N,I)O}\sim 1$)\cite{pilaf}.
Now using Eq.\ref{noeps}, the total CP asymmetry parameters in the quasi-degenerate limit of the RH masses can be written as 
\bea
(\varepsilon_1+\varepsilon_2)^{\rm NO}=\frac{(m_3-m_2)M}{8\pi v^2 \delta_{\rm lep}^{\rm NO}}\sqrt{1-\left[\frac{M^3(m_3+m_2)}{\Delta M_H^2 8\pi^2 v^2}\right]^2}, \label{qdep1}\\
(\varepsilon_1+\varepsilon_2)^{\rm IO}=\frac{(m_2-m_1)M}{8\pi v^2 \delta_{\rm lep}^{\rm IO}}\sqrt{1-\left[\frac{M^3(m_2+m_1)}{\Delta M_H^2 8\pi^2 v^2}\right]^2},\label{qdep2}
\eea
\begin{figure}
\includegraphics[scale=.35]{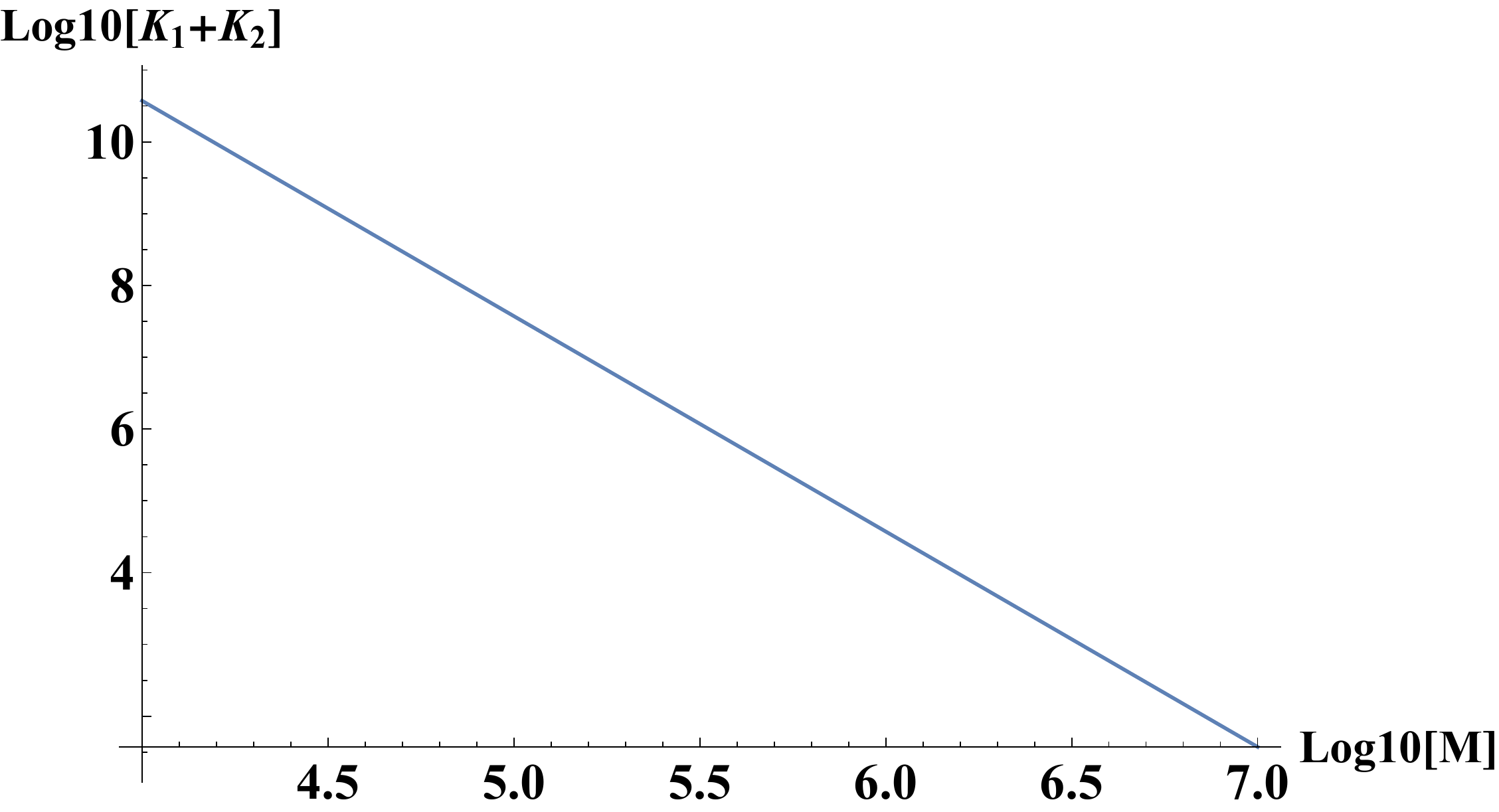}\includegraphics[scale=.35]{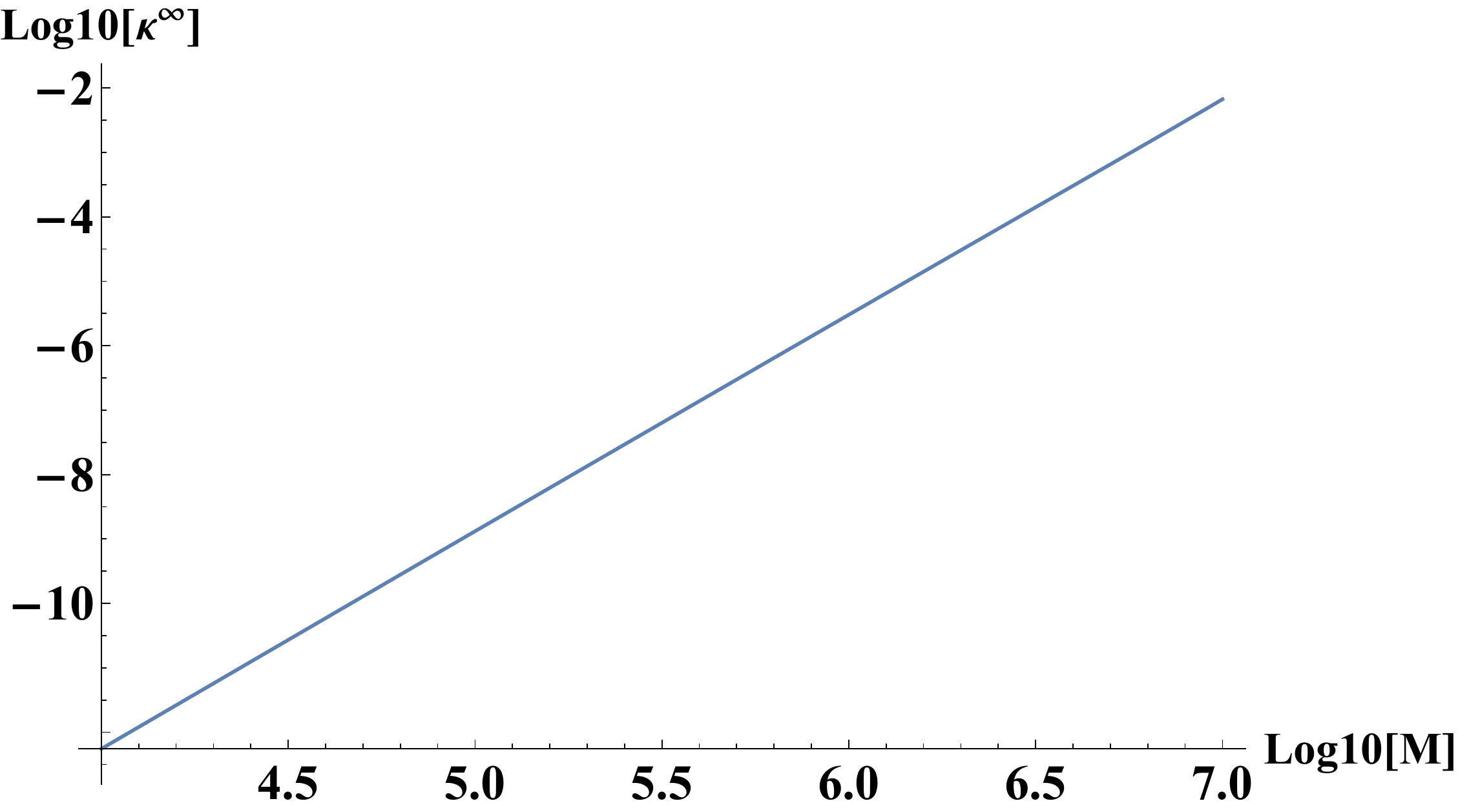}
\caption{Left: Variation of the sum of the decay parameters (which is responsible for washout of thermally generated lepton asymmetry) with the RH mass scale $M$. The nature of this variation is completely due to the conditions imposed by Neutrino Option. Right: variation of the efficiency factor $\kappa^\infty$ with $M$.}\label{decay_para}
\end{figure}
where $M$ is the overall mass scale of the RH neutrinos and we ignore the regulator term in the CP asymmetry parameters. Using the master formula presented in Eq.\ref{master}, it is now easy to find the values of the $R^{\rm (N,I)O}$ which satisfy correct baryon asymmetry and correct DM relic density. In the upper panel of Fig.\ref{pspace}, we show the allowed values of the $R$ parameters in $M-M_{\rm DM}$ plane. The green region in each plot corresponds to a super degenerate case ($R^{\rm (N,I)O}<1$) \cite{Dev:2017wwc} which we do not study in this paper. The gray shades are ruled out due the similar constraint derived in Eq.\ref{con2} with the maximum branching ratio corresponding to free streaming half mode $k_{\rm FS}^{1/2}\sim 40 {\rm h}/{\rm Mpc }$\cite{Boulebnane:2017fxw} for $m_\phi=10^7$ GeV and $m_\phi=10^8$ GeV respectively $-$ we take the former value of $m_\phi$ to be the minimum one so that we can cover the maximum parameter space, i.e., $m_\phi>M^{\rm (N,I)O}_{\rm max}+M_{\rm DM}$. As mentioned earlier in the introduction, to draw the gray shaded region we use a more accurate value of the reheat temperature by generalising the condition $M_i>T_{\rm RH}$ to $M_i>\Phi(K_i)T_{\rm RH}$, where the function $\Phi(K_i)$ determines the value of $z_f=M/T$ after which washout processes go out of equilibrium. An explicit expression of the function can be calculated analytically\cite{bari} and for quasi degenerate heavy neutrinos it is given by 
\bea
\Phi\left(\sum_i K_i\right)=2+4 \left(\sum_i K_i\right)^{0.13}+e^{\frac{-2.5}{\sum_i K_i}},
\eea
where from  Eq.\ref{thre0}, $\sum_iK_i$ can be expressed as a function of the Higgs and the RH neutrino masses as
\bea
K_1+K_2=\frac{8\Delta M_{ H}^2\pi^2v^2}{m^* M^3}.\label{wo}
\eea 
  The blue dotted lines correspond to the free streaming lengths $\lambda_{\rm FS}\simeq 10 $ kpc and $\lambda_{\rm FS}\simeq 1 $ kpc for $m_\phi =10^7$ GeV, where the former separates the CDM and WDM regions\cite{fs1,fs2} and we draw the latter to indicate that  $\lambda_{\rm FS}$ decreases as DM mass increases$-$ indicating a CDM region. Note that in both the plots, there exists a small range of the RH mass scale (from the point where the $\lambda_{\rm FS}=10$ kpc curve and $R^{\rm (N,I)O}=1$ intersect to the point where curves correspond to $m_\phi=10^7$ GeV and $R^{\rm (N,I)O}=1$ intersect) in which the parameter space favours only WDM. We would also like to mention that the lower bounds on the RH masses we obtain in this case ($M^{\rm NO}>1.1\times 10^6$ GeV and $M^{\rm IO}>2.6\times 10^6$) are more or less similar to those obtained in the thermal case ($M^{\rm NO}>1.2\times 10^6$ GeV and $M^{\rm IO}>2.4\times 10^6$)\cite{lepNeo1}.  It is evident that for NO, there is a large parameter space which allows mild-resonant solutions ($R^{\rm NO}\gg 1$). However, for IO, the allowed parameter space gets reduced. This is since, compared to the NO,  CP asymmetry parameter is suppressed by the pre-factor $m_2-m_1$ (cf. Eq.\ref{qdep2}) in case of IO. Now we turn to case of pair production, i.e., $\phi\rightarrow N_i N_i$. In the lower panel, we show the allowed parameter space in the $\sum_i\varepsilon_i-M$ plane. The gray shades are due to the constraints derived from Eq.\ref{bau} and it reads
  \bea
  \varepsilon_1+\varepsilon_2\geqslant N_{B-L}^{\rm Obs} \left(\frac{15 m_\phi}{\pi^4 B_\chi g_{\rm eff}}\right)\frac{\Phi (\sum_i K_i)}{M}.
  \eea
  
Taking $B_\chi\sim10^{-2}$,  compare to the thermal leptogenesis, we get more relaxed lower bounds on the RH masses for both the  ordering of light neutrino masses as
\bea
M^{\rm NO}\gtrsim 2.6\times 10^5 {\rm GeV},~~M^{\rm IO}\gtrsim 6\times 10^5 {\rm GeV}.\label{mmax}
\eea
Consequently the upper bounds on the reheat temperature come out as
\bea
T_{\rm RH}^{\rm NO}\leq 9.1\times 10^3 {\rm GeV},~~T_{\rm RH}^{\rm IO}\leq 2.8\times 10^4 {\rm GeV}.
\eea
Note that for NO, the lower bound is almost an order of magnitude below what is obtained in the thermal case. These bounds can be even more relaxed for larger values of $B_\chi$\cite{nth6,shafi}.\\

One might wonder why the mild-resonant solutions appear in our case, while for the thermal production of right-handed neutrinos, its clearly resonant  even if the RH mass scale is $\sim 10^6$ GeV, as confirmed by Ref.\cite{lepNeo1}. The basic physical reason is, in the thermal case, Neutrino Option introduces a rapid exponential reduction of production efficiency with the decrease of RH mass scale.  This can be seen by calculating the efficiency factor, say e.g., 
\bea
\kappa_i (z=M_{1}/T)=-\int_{z_{\rm T_{RH}}}^z \frac{dN_{N_i}}{dz^\prime}e^{-\sum_{i}\int_{z^\prime}^z W_{i}^{\rm ID}(z^{\prime\prime})dz^{\prime\prime}}dz^\prime\, ,\label{effi1}
\eea
where we include only inverse decays at the washout for simplicity. After properly subtracting the real intermediate state contribution of $\Delta L=2$ process\cite{pilaf,bari2}, $W^{\rm ID}$ can be written as
\bea
W_i^{\rm ID}=\frac{1}{4}K_i(1+\delta_{\rm lep})\mathcal{K}_1(z_i)z_i^3\,,\label{inv_decay}
\eea
where $z_i=z (1+\delta_{\rm lep})$ and $\mathcal{K}_1(z_i)$ is modified Bessel function. Clearly, in the limit $\delta_{\rm lep}\ll 1$, the total washout due to both the RH neutrinos can be approximated as
\bea
W_{\rm tot}^{\rm ID} \simeq \frac{1}{4}(K_1+K_2)\mathcal{K}_1(z)z^3\,.\label{inv_decay}
\eea
Thus in the quasi-degenerate limit of the RH neutrinos, the decay parameters ($K_i$) leave an additive contribution to the exponential washout. Eq.\ref{wo} justifies why even with a small decrease in $M$, the sum of the decay parameter rapidly increases (cf. Fig.\ref{decay_para}) and  strongly washes out the produced lepton asymmetry. Thus one needs a resonant enhancement in the CP asymmetry parameter to compensate this strong washout. In the strong washout regime, the Yukawas will quickly thermalise the heavy neutrinos and the final asymmetry which is independent of initial conditions can be written as
\bea
N_{B-L}^{\rm Thermal} = (\varepsilon_1 +\varepsilon_2)\kappa^\infty,
\eea
where $\kappa^\infty$ is given by\cite{bari,Samanta}
\bea
\kappa^\infty =\frac{2}{\sum_i K_i \Phi(\sum_i K_i)}(1-e^{\frac{\sum_i K_ i  \Phi(\sum_i K_i}{2}}).
\eea
In the right panel if Fig.\ref{decay_para}, we plot the efficiency factor with the mass  scale $M$. Notice that around $M\sim 1.2\times10^6$ GeV, $\kappa^\infty\sim 6\times 10^{-6}$  which requires $\varepsilon_1+\varepsilon_2\sim 10^{-2}$ to be consistent with $N_{B-L}^{\rm Obs}$. Now going back to the  bottom left panel of Fig.\ref{pspace}, apparently, it seems that for $M\sim 1.2\times10^6$ GeV, the value of  $\varepsilon_1+\varepsilon_2\sim 10^{-2}$ is out side the parameter space (i.e., the value is  in the green region). However, recall that in thermal leptogenesis flavour effects which reduces the strength of the washout and increases the production efficiency\cite{fl1,fl2,fl3,fl4}, plays an important role and thus one can have the required CP asymmetry within the allowed parameter space. In any case, the rough estimate presented above clearly shows why around $M\sim 10^6$ GeV, the CP asymmetry parameter tends to hit $R\sim 1$ contour (resonant solution) for thermal leptogenesis.
On the other hand, as already mentioned, for $\phi\rightarrow N_i\chi$ decays, we obtain approximately similar (what is obtained for thermal leptogenesis) allowed range for the RH neutrinos masses. Thus around $M\sim 10^6$ one should get $R\sim 1$ solutions which is exactly the case as shown in the upper left panel of Fig.\ref{decay_para}.  In this case, though the efficiency factor 
\bea
\kappa_{\rm Dark}=N_{B-L}^{\rm Obs }\left(\frac{\rm GeV}{1.22 M_{\rm DM}}\right)
\eea
is not constrained by the Neutrino Option,  Eq.\ref{con2} with $M_i>\Phi(K_i)T_{\rm RH}$ for which we draw the blue and the gray shades,  is affected by the Neutrino Option.
\begin{figure}
\includegraphics[scale=.6]{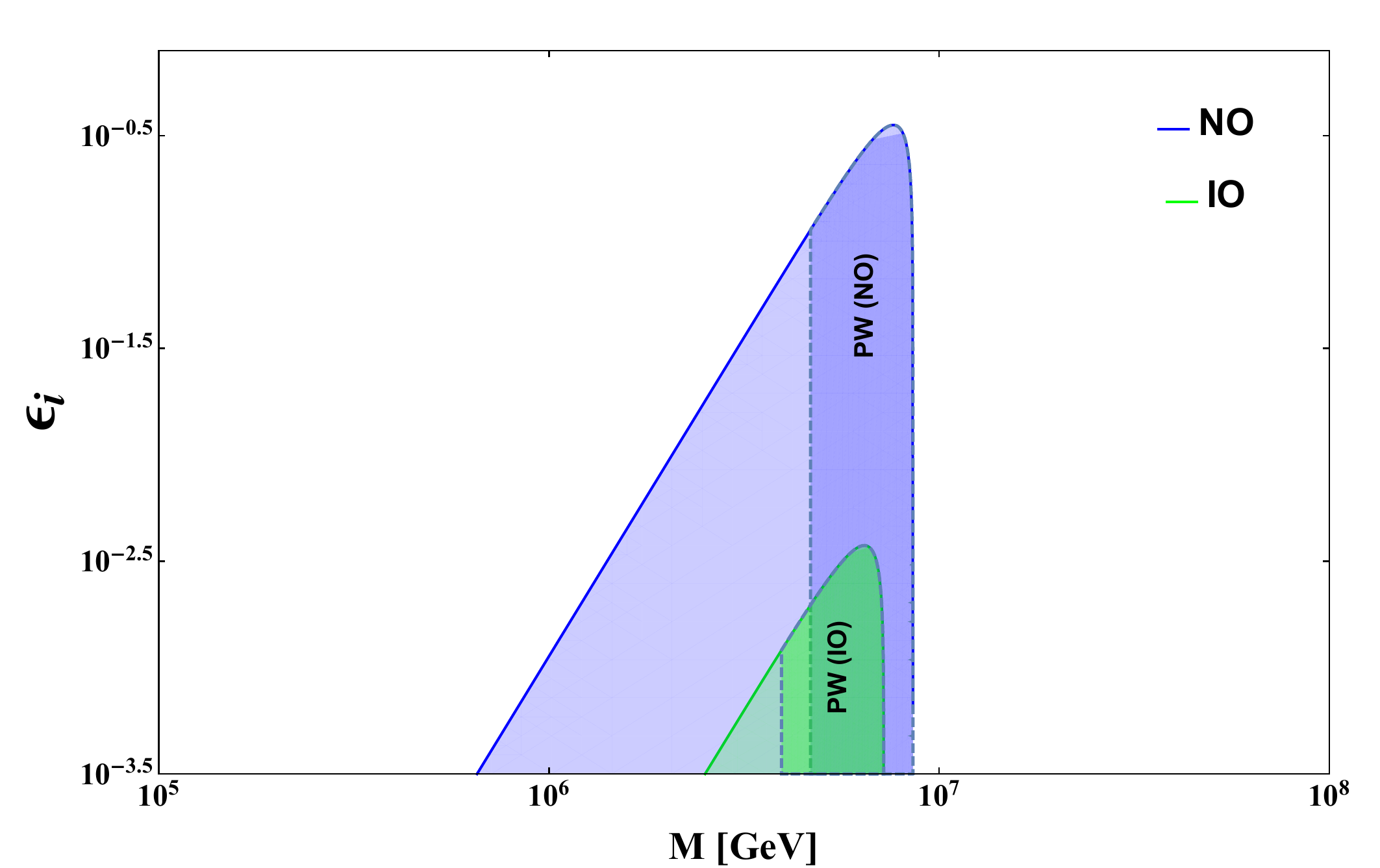}
\caption{Variation of the total CP asymmetry parameter with the RH mass scale $M$ allowing Pilaftsis-Underwood resonance. The parameter space plotted in purple (green) is for normal mass ordering (inverted mass ordering). Maximum enhancement in the CP asymmetry parameter occurs at $M_{\rm max}=7.63\times10^6$ GeV ($M_{\rm max}=6.43\times10^6$ GeV) for normal mass ordering (inverted mass ordering).}\label{cpmax}
\end{figure}

 We conclude with the following remarks:
 
I) Since we have discussed quasi-degenerate RH mass spectrum and hence a single scale seesaw EFT\cite{rg1,rg2}, we do not expect significant affect of RG group evolution on the derived parameter space.

II) Fig.~\ref{fig1} shows that the typical range of $m_{\phi}$ in such a process is $10^8-10^{12}{~\rm GeV}$, which is in contradiction with the observational constraints on simple single field models of inflation. As an example, the mass of the inflaton in a chaotic quadratic inflation with $m_{\phi}^2\phi ^2$ potential is constrained from the precisely measured amplitude of the primordial scalar power spectrum in Planck 2018~\cite{Akrami:2018odb} to be $m_{\phi} \sim 10^{13}{~\rm GeV}$. However, multi-field models of inflation where the mass of the inflaton does not determine cosmological observables in Planck, can accommodate such values of $m_{\phi}$ as required in the DM and neutrino productions in this context. For example, hybrid inflation models~\cite{Linde:1993cn,GarciaBellido:1997wm} can be good candidates for this case where the waterfall criteria are satisfied by tuning the mass and couplings of the waterfall field accordingly. However, when one analyses particular inflation models conforming with such a scenario, the bounds on the reheating temperature are also to be considered accordingly since the resulting parameter ranges for $M_{\rm DM}$ and $M$ depend on $T_{\rm RH}$. We have neglected also, e.g., possible impact of the strong Higgs-inflaton coupling which might put additional constraints on the parameter space. In any case, NeO itself requires a non-trivial conformal UV theory and so is the present scenario. The purpose of the entire study was to emphasise the difference between thermal and non-thermal production of lepton asymmetry using the constraints that come out of the NeO. \\
 
 III)  The  lower bounds on $M$ in Eq.\ref{mmax} are obtained for  $\Delta M\simeq \Gamma_i$. In this regime, there could be another contribution to the CP asymmetry parameter due the RH neutrino oscillation in the thermal bath\cite{dev1,dev2}. In this case the loop function $g(x_{ij})$ in Eq.\ref{ncp} can be replaced with
 \bea
 g(x_{ij})^{\rm tot}\equiv g(x_{ij})^{\rm mix}+g(x_{ij})^{\rm osc},
 \eea
 where $g(x_{ij})^{\rm mix}$ is the usual loop function given in Eq.\ref{gxij} and $g(x_{ij})^{\rm osc}$ is given by
 \bea
 g(x_{ij})^{\rm osc}=\frac{(1-x_{ij})\sqrt{x_{ij}}}{(1-x_{ij})^2+M_i^{-2}(\Gamma_i +\sqrt{x_{ij}} \Gamma_j )^2\frac{\rm det [Re(f^\dagger f)]}{(f^\dagger f)_{ii}(f^\dagger f)_{jj}}}.
 \eea
When the effect of  $g(x_{ij})^{\rm osc}$ is taken into account, the lower bounds on $M$ relax to
\bea
M^{\rm NO}\gtrsim 2.1\times 10^5 {\rm GeV},~~M^{\rm IO}\gtrsim 5.3\times 10^5 {\rm GeV}.
\eea

IV) For a particular value of $x$ in the orthogonal matrix, there is a global maximum in the CP asymmetry parameter. In Fig.\ref{cpmax}, we show a representative parameter space in the $\varepsilon_i-M$ plane. Thus there is a distinct heavy RH scale for leptogenesis which can be found by obtaining the value of $M$ at $\varepsilon_{\rm max}$. Maximizing the CP asymmetry parameter w.r.t $x$, we calculate them to be
\bea
M^{\rm NO}\simeq 7.63\times 10^6 {\rm GeV},~~M^{\rm IO}\simeq 6.43\times 10^6 {\rm GeV}.\label{mmax}
\eea
Otherwise, there is always a pair of $M$ for obtaining $\varepsilon_i$ of same magnitude (see Fig.\ref{cpmax}). Note that, this  typical feature of $\varepsilon_i$ is driven by the newly introduced the phantom window which is a consequence of the consistency relation derived in Eq.\ref{thre}. 
 To distinguish such a mass degeneracy one needs to dig the dynamical origin of the RH masses in a UV theory\cite{B} and study its signatures such as Gravitational wave which we shall consider in a future work.
\section{summary }\label{s4}
We discuss a non-thermal production mechanism of RH neutrinos and Dark Matter within the framework  of Neutrino Option. Using the Type-I seesaw Lagrangian, the Neutrino Option$-$a mechanism of generating Higgs mass by the quantum effects of the RH neutrinos, puts an upper bound on the RH mass scale as $M\lesssim10^7$ GeV and thereby does not facilitate hierarchical thermal leptogenesis. The parameter space for thermal resonant leptogenesis ($\Delta M\sim \Gamma_i$) is highly constrained due an increasing strong washout of the produced asymmetry with the decrease of the RH mass scale. In this article, we show that non-thermal pair production of the RH neutrinos from inflaton decay allows the RH mass scale to be smaller by more than an order of magnitude than what is obtained for the thermal case. Specially for normal light neutrino mass ordering, which is now favoured by oscillation experiments, the parameter space is in general mildly resonant ($\Delta M\gg \Gamma_i$).  Then we show that the scenario of simultaneous production of RH neutrinos and a Dark Matter is as constrained as the thermal resonant leptogenesis. The primary restriction on the parameter space in this case comes from the branching fraction of inflaton decay to RH neutrinos and Dark matter. The branching fraction is bounded from above by Ly$\alpha$ constraint on Dark Matter free streaming. The maximum accessible Dark Matter mass in this scenario is approximately 320 MeV. Depending upon the inflaton mass and branching fraction, parameter spaces for Cold as well as Warm Dark Matter can also be separated. Finally, we show that in non-thermal leptogenesis scenario the Neutrino option introduces a ``phantom window" in which the CP asymmetry parameter responsible for leptogenesis decreases with the increase of RH mass scale and minimally fine-tuned seesaw models naturally exhibit this phantom window.
\section*{Acknowledgement}
RS is supported by a Newton International Fellowship (NIF 171202). SB is supported by institute postdoctoral fellowship from Physical Research Laboratory. The authors would like to acknowledge the hospitality of WHEPP-XVI held at IIT Guwahati, INDIA, where this project was initiated.
\appendix
\section{Production of $\chi$ from scattering of RH neutrinos}
Apart from the decay of inflaton, $\chi$ can also
be produced from the scattering of RH neutrinos. In this
section we have estimated the contribution of this production
channel to the relic abundance of $\chi$. Following the prescription
given in \cite{Hall:2009bx}, we can write the collision term resulting
from the scatterings $N_i N_j \rightarrow \bar{\chi} \chi$ as follows
\begin{eqnarray}
\mathcal{C}_{N_i N_j \rightarrow \bar{\chi} \chi} = 
\dfrac{T}{1024\,\pi^6} \sum_{i,\,j=1}^2\,\int^\infty_{(M_{N_i} + M_{N_j})^2}
d\hat{s}\,\int d\Omega\,P_{N_i\,N_j}P_{\bar{\chi}\,\chi}
\left|\mathcal{M}\right|^2_{N_i N_j \rightarrow \bar{\chi} \chi}
\dfrac{K_1\left(\frac{\sqrt{\hat{s}}}{T}\right)}{\sqrt{\hat{s}}}\,,
\label{CE}
\end{eqnarray}
where $\hat{s}$ and $P_{\alpha \beta}$ are total energy and
magnitude of 3-momentum with respect to the centre of momentum frame
for initial and final states respectively. The momentum $P_{\alpha \beta}$
has the following expression
\begin{eqnarray}
P_{\alpha \beta} = \dfrac{\sqrt{\hat{s} - (m_{\alpha} + m_{\beta})^2}
\sqrt{\hat{s} - (m_{\alpha} - m_{\beta})^2}}{2\,\sqrt{\hat{s}}}\,,
\end{eqnarray}  
with $m_{\alpha}$, $m_{\beta}$ are the masses of particles $\alpha$
and $\beta$ in either initial or final states. Furthermore, $d\Omega$ is the
solid angle subtended by one of the outgoing particles with
respect to an arbitrary direction which can be considered the
direction of an incoming particle. $K_1(\frac{\sqrt{s}}{T})$ is the modified Bessel
function of second kind. The Lorentz invariant matrix
amplitude square for the scattering process ${N_i N_j \rightarrow \bar{\chi} \chi}$
is denoted by $\left|\mathcal{M}\right|^2_{N_i N_j \rightarrow \bar{\chi} \chi}$.
In the present case, the expression of matrix amplitude square is given by
\begin{eqnarray}
\left|\mathcal{M}\right|^2_{N_i N_j \rightarrow \bar{\chi} \chi} =
\dfrac{4\,y^2_{i}\,y^2_{j}}{(\hat{t}-m^2_{\phi})^2}[(M_{\rm DM}+M_i)^2-\hat{t}]
[(M_{\rm DM}+M_j)^2-\hat{t}]\,. 
\end{eqnarray}
In the above expression $\hat{t}$ is one of the Mandelstam variables and $y_i$ is the Yukawa
coupling corresponding to the interaction $\bar{\chi}\phi N_i$. The Boltzmann equation
involving the collision term $\mathcal{C}_{N_i N_j \rightarrow \bar{\chi} \chi}$ is
given by,
\begin{eqnarray}
\dfrac{dY_{\rm DM}}{dT} = - \dfrac{\mathcal{C}_{N_i N_j \rightarrow \bar{\chi} \chi}}{s\,H\,T} \,.
\end{eqnarray}
Where $Y_{\rm DM}$ is the comoving number density of $\chi$ and it is defined
as $n_{\rm DM}/s$ with $s$ being the entropy density of the universe. The Hubble
parameter is indicated by $H$.  Now, we know
that $N_{\rm DM}/Y_{\rm DM} = \dfrac{\pi^4}{45}\,g_{*s}$, where $N_{\chi}$, as defined
in the Section \ref{s2}, is the number density of
$\chi$ per $N_i$ in ultra-relativistic thermal equilibrium. Therefore, the
contribution to $N_{\chi}$ at the present epoch ($T=T_0$), due to the production
of $\chi$ from the scatterings of RH neutrinos, is 
\begin{eqnarray}
N_{\rm DM}(T_0) = - \dfrac{~\pi^4}{45}\,g_{*s}(T_0) \int^{T_0}_{T_{RH}} dT\,
\dfrac{\mathcal{C}_{N_i N_j \rightarrow \bar{\chi} \chi}}{s\,H\,T}\,,
\label{nchiT0}
\end{eqnarray}
and $T_0$ being the present temperature of the universe.

Finally, the contribution of these scattering processes to the
relic density of $\chi$ can be obtained from the following expression,
\begin{eqnarray}
\Omega_{\rm DM} h^2 = 3.2558 \times 10^7 
\left(\dfrac{M_{\rm DM}}{\rm GeV}\right)\,N_{\rm DM}(T_0)\,.
\label{omegachi}
\end{eqnarray}
To estimate how much impact these scatterings have on the relic
abundance of $\chi$, we have considered a benchmark point:
$m_{\chi}=10^{-3}$ GeV,  $M_{1}=M_{2}=M=5\times 10^6$ GeV and  
$m_{\phi}= 10^8$ GeV. For $B_\chi=10^{-4}$, the couplings $y_1=y_2=y$ can be approximated  as 
\bea
y\approx \sqrt{ \frac{16 \pi\times 10^{-4}}{m_\phi}\left(\frac{\pi^2 g^*}{10}\right)\frac{M^2}{M_{\rm Pl} \Phi\left(\sum_i K_i\right)^2}}\approx 10^{-9},
\eea
where we have taken  $T_{\rm RH}=M/\Phi\left(\sum_i K_i\right)=2.2\times 10^4$ GeV. 
Using Eqs. (\ref{nchiT0}, \ref{omegachi}), we find that the contribution to $\Omega_{\rm DM} h^2$
for this benchmark point is practically nil. This smallness is primarily
due to the chosen value of $T_{\rm RH} \ll M_{i}$, when the number densities of RH
neutrinos are  already Boltzmann suppressed. Besides, the small values of Yukawa couplings
and the large inflaton mass have greatly reduced the annihilation cross section
of $N_i\,N_j\rightarrow \bar{\chi}\chi$. Thus, we have neglected these
production processes of $\chi$ in our main discussion. 
{}
\end{document}